\documentclass[reprint, 
nofootinbib,
amsmath,amssymb,
aps,prd
]{revtex4-2}

\usepackage{graphicx}
\usepackage{bm}
\usepackage{booktabs}
\usepackage{amsmath}
\usepackage{mathrsfs}
\usepackage{subfigure}

\usepackage[hidelinks]{hyperref}
\hypersetup{
	colorlinks=true,
	linkcolor=blue,
	filecolor=gray,      
	urlcolor=blue,
	citecolor=blue,
}

\allowdisplaybreaks[1]
\begin{document}
	\title{Post-Newtonian Roche-Lobe-Overflow Prescription for Compact Binary Mass Transfer and the Corresponding Gravitational Waveforms}
	
	\author{Shuai Zhang$^{1}$}
	\email{zhangshuai21@lzu.edu.cn}
	
	\author{Jie Yang$^{1,2,3,4}$}
	\email[Contact author: ]{yangjiev@lzu.edu.cn}

	\author{Zi-Han Zhang$^{5,6}$}
	
	\author{Shenghua Yu$^{7}$}

	\affiliation{$^{1}$School of Physical Science and Technology, Lanzhou University,Lanzhou 730000, China }
	
	\affiliation{$^{2}$ Institute of Theoretical Physics \& Research Center of Gravitation, Lanzhou University, Lanzhou 730000, China}
	
	\affiliation{$^{3}$Key Laboratory of Quantum Theory and Applications of MoE, Lanzhou University, Lanzhou 730000, China}
	
	\affiliation{$^{4}$Lanzhou Center for Theoretical Physics \& Key Laboratory of Theoretical Physics of Gansu Province, Lanzhou University, Lanzhou 730000, China}
	
	\affiliation{$^{5}$International Centre for Theoretical Physics Asia-Pacific, University of Chinese Academy of Sciences, 100190 Beijing, China}
	
	\affiliation{$^{6}$Taiji Laboratory for GW Universe, University of Chinese Academy of Sciences, Beijing 100049, China}
	
	\affiliation{$^{7}$ CAS Key Laboratory of FAST, National Astronomical Observatories, Chinese Academy of Sciences, 20A Datun Road, Beijing 100101, China}
	
\begin{abstract}
	Mass transfer in binary systems is central to many astrophysical phenomena, including the evolution of compact interacting binaries. Starting from the first post-Newtonian hydrodynamic equations in the corotating frame, we derive the first post-Newtonian Roche potential and construct the corresponding post-Newtonian form of the Roche lobe overflow mass transfer prescription. We then include the time dependence of the component masses in the binary dynamics and compute the associated corrections to the equations of motion, gravitational-wave energy and angular-momentum fluxes, and far-zone polarization waveforms. Finally, we apply the model to representative ultracompact binary systems. We find that mass transfer can play an important role in the dynamical evolution of compact binaries. For gravitational-wave observations, its main effect appears as a secular phase drift accumulated over long observation times.
\end{abstract}

		\maketitle	
	\section{Introduction}
	
	Binary evolution underlies many high-energy sources and gravitational-wave progenitors. X-ray binaries \cite{Done2007}, cataclysmic variables \cite{Knigge2011}, novae \cite{Chomiuk2021}, Type Ia supernova progenitors \cite{Branch1992}, millisecond pulsars \cite{Lorimer2008}, and compact binary mergers \cite{Abbott2016} are all shaped by binary interactions.
	
	Mass transfer (MT) is one of the main channels through which binary interactions affect stellar evolution. It was first introduced to explain the Algol paradox and the spectral variability of interacting eclipsing binaries \cite{Crawford1955,Morton1960,Giuricin1983}. It is now a standard ingredient in models of X-ray binaries \cite{Heuvel1973}, cataclysmic variables \cite{Marsh2004}, nova outbursts \cite{MacDonald1980}, and Type Ia supernova
	progenitors \cite{Whelan1973,Trimble1982,Iben1984}.
	
	The standard description of stable MT in close binaries is based on Roche-lobe overflow (RLO). When the donor fills its Roche lobe, gas can pass through the inner Lagrangian point and flow toward the accretor \cite{Eggleton1983}. Early studies
	established this picture and its role in binary evolution \cite{Paczynski1966,Kippenhahn1967,Plavec1967}. Later work included the hydrodynamic structure, thermodynamic state, and optically thin or optically thick character of the stream \cite{Lubow1975,Ritter1988,Verbunt1988,Kolb1990}. These developments improved quantitative estimates of the MT rate, and recent work has revisited the local RLO nozzle and stream geometry in a more general form \cite{Cehula2023,Jackson2017}.
	
	For compact interacting binaries, relativistic corrections can become relevant. The post-Newtonian (PN) approximation provides a systematic description of the weak-field and slow-motion regime of general relativity \cite{Blanchet1995,Blanchet2024}. It is widely used in compact-binary dynamics and gravitational-wave modeling. In ultracompact X-ray binaries (UCXBs) \cite{Nelemans2010,Yu2021}, the orbital separation is small. The effective potential and orbital velocity can therefore receive non-negligible PN corrections. This is important for RLO, because the MT rate is highly sensitive to the effective potential difference between the donor photosphere and the
	inner Lagrangian point \cite{Lubow1975,Ritter1988,Cehula2023}. Recent studies have developed mass-transfer corrections to compact-binary waveforms, orbital dynamics, detectability, and cosmological-redshift measurements \cite{Zhang2024,Zhang20251,Zhang20252,Zhang20262,Yang2025}. Related PN extensions have also been studied for charged compact binaries \cite{Zhang20261}.
	
	Standard PN treatments of compact-binary dynamics usually take the component masses as constant parameters. By contrast, standard RLO prescriptions are built in a Newtonian Roche geometry. This separation is not fully satisfactory
	for mass-transferring compact binaries. A PN description of such systems should include both the correction to the matter stream near \(L_1\) and the effect of time-dependent masses on the orbital dynamics and gravitational radiation.
	
	In this work we construct a 1PN model of RLO. The compact objects are treated as point masses, while the transferred stream is treated as a fluid moving in the binary gravitational field. Starting from the 1PN hydrodynamic equations in the corotating frame, we derive the 1PN Roche potential from the Bernoulli relation. We also recover the same potential from the relativistic effective potential constructed with the helical Killing vector of a conservative quasi-circular binary \cite{Gourgoulhon2001,Friedman2002}. We then use
	this potential to obtain the PN form of the RLO mass-transfer prescription.
	
	The paper is organized as follows. In Sec.~II, we derive the 1PN Roche potential and the RLO prescriptions. In Sec.~III, we introduce time-dependent masses into the binary equations of motion. In Sec.~IV, we compute the MT corrections to the gravitational-wave energy and angular-momentum fluxes. In Sec.~V, we derive the corresponding corrections to the far-zone polarization waveforms. In
	Sec.~VI, we present numerical results for representative binaries and discuss the application to 4U 1728-34.
   
	\section{ PN Mass Transfer Model}
	\subsection{Post-Newtonian Hydrodynamics}
	
	We start from the post-Newtonian formulation of relativistic hydrodynamics developed by Chandrasekhar \cite{Chandrasekhar1965}. This framework describes fluids in the weak-field and slow-motion regime of general relativity. It is therefore suitable for treating the transferred gas as a fluid moving in the gravitational field of a semi-detached binary.
	
	We adopt the 1PN metric \cite{Blanchet2024,Poisson2014}
	\begin{align}
		g_{00} =& - 1 + \frac{2}{c^2}V - \frac{2}{c^4}V^2 + \mathcal{O} \left( \frac{1}{c^{6}} \right) , \\
		g_{0i} =& - \frac{4}{c^3} V_i  + \mathcal{O} \left( \frac{1}{c^5} \right) , \\
		g_{ij} =& \delta_{ij} \left[ 1 + \frac{2}{c^2} V  \right]  + \mathcal{O} \left( \frac{1}{c^4} \right).
	\end{align}
	Here \(V\) and \(V_i\) are the scalar and vector PN potentials. They are determined by
	\begin{align}
		V = \square_{\text{ret}}^{-1}[- 4 \pi G \sigma], \qquad
		V_i = \square_{\text{ret}}^{-1}[- 4 \pi G \sigma_i],
	\end{align}
	where \(G\) is the gravitational constant. The source densities are defined by \cite{Blanchet2024}
	\begin{align}
		\sigma = \frac{1}{c^2}(T^{00} + T^{ii}), \qquad
		\sigma_i = \frac{1}{c}T^{0i}.
	\end{align}
	
	For the fluid component, we write the energy-momentum tensor as
	\begin{align}
		T^{\mu\nu}_{\mathcal{F}}
		= \left(\rho + \epsilon/c^2 + p/c^2\right) u^\mu u^\nu
		+ p g^{\mu\nu}.
	\end{align}
	Here \(\rho\), \(\epsilon\), and \(p\) denote the rest-mass density, internal-energy density, and pressure of the fluid, respectively. The four-velocity is denoted by \(u^\mu\). In the hydrodynamic equations below, \(U\) denotes the Newtonian part of the scalar potential \(V\).
	
	At 1PN order, the mass-conservation law and Euler equation are \cite{Chandrasekhar1965,Poisson2014}
	\begin{align}
		\partial_t \rho^* &+ \partial_i( \rho v^i) = 0, \label{Eq.Conservation} \\ 
		\begin{split}
		\rho^* \frac{d v^j}{dt}
		&= -\partial_j p + \rho^* \partial_j V \\
		& + \frac{1}{c^2} \left[
		\left( \frac{1}{2} v^2 + V + \Pi + \frac{p}{\rho^*} \right) \partial_j p - v^j \partial_t p
		\right] \\
		& + \frac{1}{c^2} \rho^* \left[
		(v^2 - 4V)\partial_j V - v^j \left(3\partial_t V + 4 v^k \partial_k V\right) \right. \\
		& \left.
		+ 4\partial_t V_j + 4 v^k \left(\partial_k V_j - \partial_j V_k\right) 
		\right]  + \mathcal{O}(c^{-4}) .
		\end{split} \label{Eq.Euler}
	\end{align}
	Here \(\rho^* = \sqrt{-g} \gamma \rho\), \(\gamma = dt/d\tau\), and \(v^i\) is the coordinate velocity of the fluid element. 
	
	\subsection{Post-Newtonian Roche Potential}
	
	Before deriving the mass-transfer rate, we specify the binary model used in this section. We treat the two stellar components as point masses and neglect the self-gravity of the transferred stream. The stream is therefore modeled as a fluid moving in the gravitational field generated by the binary.
	
	With this approximation, the PN potentials take the point-mass form \cite{Blanchet2006,Blanchet2024}
	\begin{align}
		V &= \frac{Gm_1}{r_1} + \frac{Gm_1}{c^2} \biggl[ -\frac{(n_1v_1)^2}{2r_1} + \frac{2v_1^2}{r_1} \notag  \\
		&+ Gm_2 \biggl( -\frac{r_1}{4r_{12}^2} - \frac{5}{4r_1r_{12}  }+ \frac{r_2^2}{4r_1r_{12}^3}\biggl)\biggl] + 1 \leftrightarrow 2,  \\
		V_i & = \frac{Gm_1 v_1^i}{r_1} + 1 \leftrightarrow 2,
	\end{align}
	where \(r_1=|\mathbf{x}-\mathbf{x}_1|\), \(r_{12}=|\mathbf{x}_1-\mathbf{x}_2|\), and \(\mathbf{n}_1=(\mathbf{x}-\mathbf{x}_1)/r_1\). The velocity of body \(A\) is denoted by \(\mathbf{v}_A\).
	
	To close Eqs.~\eqref{Eq.Conservation} and \eqref{Eq.Euler}, we adopt an isothermal ideal-gas equation of state, following Lubow and Shu \cite{Lubow1975},
	\begin{align}
		p = \frac{k_B}{\bar{m}} T \rho = c_T^2 \rho ,\quad \epsilon = \frac{1}{\Gamma - 1} \frac{k_B}{\bar{m}} T =  \frac{1}{\Gamma - 1} c_T^2.  \label{Eq.EOS}
	\end{align}
	Here \(k_B\) is Boltzmann's constant, \(\bar{m}\) is the mean particle mass, and \(c_T\) is the isothermal sound speed.
	
	The roche problem is usually treated in a binary corotating coordinate system, and we follow the same approach. To describe a purely rotating coordinate system, we would use the transformation
	\begin{align}
		t' = t + \mathcal{O}(c^{-4}), \quad x'^j = R^j_k(t) x^k + \mathcal{O}(c^{-2}),
	\end{align}
	where $R^j_k(t)$ is rotation matrix, which has the following relationship with the corotating angular frequency vector of the binary system
	\begin{align}
		\omega_j(t) = \frac{1}{2} \epsilon_{jkl} \dot{R}^m_k R_{ml}.
	\end{align}
	
	It is worth noting that such a coordinate transformation is not a well-behaved PN coordinate transformation. It has the issues that it prevents the metric from asymptotically approaching the Minkowski metric at infinity and that particles may exceed the local speed of light. Nevertheless, within a certain range ($\omega r \sin \theta < c (1 - V/c^2)$ ), it can indeed be very useful \cite{Poisson2014}. 
	
	In the corotating coordinate system, the Euler equation is given by
	\begin{widetext}
	\begin{align}
		\begin{split}
			&\left( \partial_t' + v' \cdot \nabla' \right) v'^i = -\frac{1}{\rho} \partial'_i p + \partial'_i V' - \left[ \omega \times (\omega \times x') \right]^i - 2(\omega \times v')^i + (\dot{\omega} \times x')^i \\
			& + \frac{1}{c^2} \left[  \left( (v' + \omega \times x')^2 -2 U' + \Pi + \frac{P}{\rho} \right) \partial'_i \ln p - (v' + \omega \times x')^i \partial_t' \ln p \right] \\
			&+ \frac{1}{c^2} \biggl[ \biggl( (v' + \omega \times x')^2 - 4U' \biggl)\partial_i' U' + (v' + \omega\times x')^i (v'+\omega \times x')^j \partial_j U'\biggl] \\
			& +\frac{1}{c^2} \biggl[ -(v' + \omega \times x')^i (3\partial_t U' + 4v^j \partial_j U') + 4 \partial_t' U^{i\prime} + 4  (v' + \omega \times x')^j ( \partial_j U^{i \prime} - \partial_i U^{j \prime})\biggl].  \label{Eq.Euler_Corotating}
		\end{split}
	\end{align}
	\end{widetext}

	In the following, all primed quantities are understood to be evaluated in the corotating frame. We omit the primes unless they are needed for clarity.
	
	For a steady isothermal flow, taking the scalar product of Eq.~\eqref{Eq.Euler_Corotating} with the velocity gives the Bernoulli relation,
	\begin{align}
		\begin{split}
			&\vec{v} \cdot \nabla \biggl[ \frac{1}{2} v^2 - V - \frac{1}{2} \omega^2 r^2 + c_T^2 \ln \rho \\
			&+\frac{1}{c^2} \biggl( -\frac{1}{4} (v^2 - \omega^2 r^2)^2 + 2 U  ( v^2 - \omega^2 r^2)  \\
			& \qquad \quad+ 4 ( \mathbf{\omega} \times \mathbf{r} ) \cdot \vec{V}  + \frac{\Gamma}{\Gamma - 1} c_T^4 \ln \rho\biggl)  \biggl] = 0.
		\end{split}
	\end{align}
	
	We assume that, near \(L_1\), the stream velocity is mainly directed along the line connecting the donor and the accretor. Similar local nozzle approximations have been used in RLO calculations \cite{Cehula2023,Jackson2017}.
	
	After isolating the terms that do not depend on the fluid's own velocity $\mathbf{v}$ and $\rho$, the remaining contribution is determined only by the spacetime geometry and the non-inertial effects of the corotating frame. We define this contribution as the 1PN Roche potential.
	\begin{align}
		\phi^{PN} =  -\frac{1}{4}\omega^4 r^4  -2V\omega^2 r^2 + 4 ( \mathbf{\omega} \times \mathbf{r} ) \cdot \vec{V} .\label{1PNRoche1}
	\end{align}
	
    Above, we derived the 1PN Roche potential from the Bernoulli relation associated with the fluid equations of motion in the corotating non-inertial frame. To justify this definition, we now introduce the helical Killing vector of a binary on a conservative quasi-circular orbit and use it to construct the corresponding relativistic effective potential.

	We consider a conservative, quasi-circular binary and neglect radiation reaction.
	In this approximation the near-zone geometry is approximately helically
	symmetric,
	\begin{equation}
		\mathcal L_k g_{\mu\nu} \simeq 0 ,
	\end{equation}
	where \(\mathcal L_k\) denotes the Lie derivative along the vector field\(k^\mu\). The helical vector is
	\begin{equation}
		k = \partial_t + \omega \partial_\phi .
	\end{equation}
	The operators \(\partial_t\) and \(\partial_\phi\) are coordinate basis vectors,
	\begin{equation}
		\partial_t \equiv \left(\frac{\partial}{\partial t}\right),
		\qquad
		\partial_\phi \equiv \left(\frac{\partial}{\partial \phi}\right).
	\end{equation}
	The vector \(\partial_\phi\) generates rotations around the orbital angular momentum axis, which is chosen to be the \(z\)-axis. In Cartesian coordinates,
	\begin{equation}
		\partial_\phi
		=
		-y\partial_x+x\partial_y .
	\end{equation}
	$k^\mu$ becomes
	\begin{equation}
		k^\mu =	\left(1,\, -\frac{\Omega y}{c},\, \frac{\Omega x}{c},\, 0 \right).
	\end{equation}
	For the 1PN metric, the norm of \(k^\mu\) is
	\begin{align}
		-k^2=1 &-\frac{1}{c^2}\left( 2V+\omega^2r^2\right)  \notag \\
		&+\frac{1}{c^4}\left[ 2V^2 +8\omega\mathcal V_\phi -2V\omega^2 r^2\right] +\mathcal O(c^{-6}) ,
	\end{align}
	where the $\mathcal V_\phi$ is defined as
	\begin{align}
		\mathcal V_\phi=-yV_x+xV_y.
	\end{align}
	The corotating lapse factor is
	\begin{equation}
		N_k=\sqrt{-k^2}.
	\end{equation}
	We define the 1PN effective potential by
	\begin{equation}
		\phi_{\rm eff}=	c^2\ln N_k = \frac{c^2}{2}\ln(-k^2).
	\end{equation}
	Expanding the logarithm to 1PN order gives
	\begin{align}
		\begin{split}
			\phi_{\rm eff} = &-V -\frac{1}{2}\omega^2r^2   \\  
			& + \frac{1}{c^2} \left[ -\frac{1}{4}\omega^4 r^4  -2V\omega^2 r^2 +4\omega \mathcal V_\phi  \right] \label{1PNRoche2}.
		\end{split}
	\end{align}
	It is clear that Eq.~\eqref{1PNRoche2} reduces exactly to the Newtonian Roche potential in the Newtonian limit. At 1PN order, it also agrees with the Roche potential obtained from Bernoulli relation.
	\subsection{Optically-thin Mass Transfer Function}
	The stationary optically thin prescription for mass transfer was developed by Ritter \cite{Ritter1988}, building on the work of Lubow and Shu \cite{Lubow1975}. This prescription applies when the donor atmosphere extends to the inner Lagrangian point, even if the photospheric radius lies slightly inside the Roche lobe. 
	The gas above the photosphere is optically thin to radiation from the donor, but remains in thermal equilibrium with the local radiation field. The flow can be modeled as isothermal, with \(T\simeq T_{\rm eff}\), where \(T_{\rm eff}\) is the donor effective temperature. Near \(L_1\), the gas reaches the isothermal sound speed \(c_T\). This leads to the standard estimate of the mass-transfer rate in terms of the density and effective cross section of the stream at \(L_1\).
	
	The mass-transfer rate via the inner Lagrangian point $L_1$ is given by \cite{Ritter1988}
	\begin{equation}
		-\dot{m} _{\rm thin} = \rho_{L} c_T Q_p ,
	\end{equation}
	where the $Q_p$ is the effective cross-section corresponding to the gas density profile at $L_1$. 
	
	The 1PN Bernoulli relation 
		\begin{align}
		\begin{split}
			&\vec{v} \cdot \nabla \biggl[ \frac{1}{2} v^2 - V - \frac{1}{2} \omega^2 r^2 + c_T^2 \ln \rho \\
			&+\frac{1}{c^2} \biggl( -\frac{1}{4} (v^2 - \omega^2 r^2)^2 + 2 U  ( v^2 - \omega^2 r^2)  \\
			& \qquad \quad+ 4 ( \mathbf{\omega} \times \mathbf{r} ) \cdot \vec{V}  + \frac{\Gamma}{\Gamma - 1} c_T^4 \ln \rho\biggl)  \biggl] = 0.
		\end{split}
	\end{align}
	Integrating the Bernoulli relation between the donor photosphere and \(L_1\), and assuming that the gas leaves the photosphere with negligible velocity, gives
	\begin{align}
		\rho_L = \mathcal{A} \rho_{\text{ph}}  \exp \biggl[ \phi^N_1 - \phi^N_{ph} + \frac{1}{c^2} \biggl(\phi_1^{PN} - \phi_{ph}^{PN} \biggl) \biggl],
	\end{align}
	The superscripts \(N\) and \(PN\) denote the Newtonian and 1PN parts of the effective Roche potential, respectively.
	where the $\mathcal{A}$ 
	\begin{align}
		\begin{split}
		\mathcal{A} = \frac{1}{c_T^2}\exp \biggl[ -\frac{1}{2} &+  \frac{1}{c^2} \biggl( -\frac{1}{4} c_T^2 + \frac{1}{2} \omega^2 r_L^2 + 2U_L\biggl)  \\
		 &+\frac{1}{c^2} \frac{\Gamma}{\Gamma-1} \biggl(  \frac{1}{2} c_T^2 + \phi^N_1 - \phi^N_{ph} \biggl) \biggl].
		 \end{split}
	\end{align}
	
	To evaluate $Q_p$, Ritter \cite{Ritter1988} followed Meyer \& Meyer-Hofmeister \cite{Meyer1983} who calculated the hydrostatic isothermal drop-off of the density in the $L_1$-plane \cite{Cehula2023}.   In our notation, we obtain
	\begin{align}
		Q_p = \frac{2 \pi }{\sqrt{B_{\rm PN}C_{\rm PN}}} c_T^2.
	\end{align}
	where the  $B_{PN}, C_{PN} $ are givin in Appendix~ \ref{Effective cross sections}.
	\subsection{Optically-thick Mass Transfer Function}
	Kolb and Ritter \cite{Kolb1990} developed the standard prescription for stationary mass transfer in the optically thick regime. In this regime, the gas below the donor photosphere is optically thick. Heat exchange within the stream and between the stream and the ambient radiation field is neglected on the flow timescale. The flow is therefore treated as adiabatic rather than isothermal. Gas with \(\phi<\phi_1\) remains in hydrostatic equilibrium, whereas gas with \(\phi_1<\phi<\phi_{\rm ph}\) flows toward \(L_1\).  Near \(L_1\), the gas reaches the adiabatic sound speed \(c_s\). The total mass-transfer rate is then obtained by integrating the mass flux over all streamlines that cross \(L_1\).
	
	The mass flux through the $L_1$ plane is written as
	\begin{equation}
		-\dot m_{\rm thick}
		=
		\int_{L_1{\rm -plane}} \rho_L c_s \, dQ ,
	\end{equation}
	where $\rho_L$ and $c_s$ are evaluated at the point where a given
	streamline crosses the $L_1$ plane.  For an optically thick stream we
	use a polytropic equation of state along each streamline,
	\begin{equation}
		p=K\rho^\Gamma, \qquad
		c_s^2=\Gamma {p\over \rho}.
	\end{equation}
	The streamline is labelled by the value of the effective potential
	$\bar\phi$ in the hydrostatic donor envelope.  The corresponding
	hydrostatic density and temperature are denoted by
	$\bar\rho(\bar\phi)$ and $\bar T(\bar\phi)$.
	
	Combining the 1PN Bernoulli relation with the polytropic equation of
	state gives, to 1PN order,
	\begin{equation}
		\rho_L(\bar\phi) =
		\left( {2\over \Gamma+1} \right)^{1/(\Gamma-1)}
		\bar\rho(\bar\phi).
	\end{equation}
	
	Using the local quadratic expansion of the 1PN effective potential in
	the $L_1$ plane, the area element satisfies
	\begin{equation}
		\left. {dQ\over d\phi} \right|_{L_1}
		=
		{2\pi\over \sqrt{B_{PN}C_{PN}}}.
	\end{equation}
	Therefore the optically thick mass-transfer rate becomes
	\begin{align}
		-\dot m_{\rm thick} = -\dot m_{\rm thin,0}  +\frac{dQ}{d\phi} \bigg|_{L_1}
		F_3 K^{1/2} P^{\frac{3\Gamma-1}{2\Gamma}} \bigg|^{P_{ph}}_{P_1},
	\end{align}
	where
	\begin{equation}
		F_3(\Gamma) = \Gamma^{1/2} \left( {2\over \Gamma+1} \right)^{ (\Gamma+1)/[2(\Gamma-1)] }.
	\end{equation}
	The first term, $-\dot m_{\rm thin,0}$, is the saturated optically-thin
	contribution, obtained from the optically-thin expression by setting
	$\phi_{\rm ph}=\phi_1$.  The second term accounts for the additional
	optically thick layers between $\phi_1$ and $\phi_{\rm ph}$.  Thus the
	1PN correction enters through both the position and curvature of the
	$L_1$ throat, encoded in $B_{PN}$ and $C_{PN}$, and through the 1PN
	Bernoulli correction along each streamline.
	\section{Equation of Motion with Mass Transfer}
	\subsection{Equation of Motion}
	
	In the standard PN treatment of compact binaries, the component masses are
	usually assumed to be constant. For a mass-transferring binary, this assumption
	must be relaxed because the exchange of matter changes the particle part of the
	stress-energy tensor and feeds back on the orbital motion. Variable-mass
	two-body dynamics is known to require care in defining the effective force and
	momentum balance \cite{Hadjidemetriou1963,Plastino1992}. We model the two
	compact objects as point particles with slowly varying masses. The particle
	contribution to the stress-energy tensor is written schematically as
	\begin{equation}
		T^{\mu\nu}
		=
		m_1(t)
		\frac{d x^\mu}{ds}
		\frac{d x^\nu}{ds}
		\delta(x-y_1)
		+1\leftrightarrow 2 .
	\end{equation}
	
	The total stress-energy tensor of the particles and the transferred stream is
	conserved. If the stream contribution is encoded in time-dependent particle
	masses, the divergence of the particle part produces an effective
	mass-transfer term. 
	
	From the conservation of the energy-momentum tensor
	\begin{align}
		\begin{split}
			\nabla_\nu T^{\mu\nu} &= \nabla_\nu \left[m_1(t) \frac{d x^\mu}{ds} \frac{dx^\nu}{ds} \delta(x-y_1) + 1 \leftrightarrow 2 \right] \\
			&=\nabla_\nu \left( \tilde{m} _1\frac{dx^\nu}{ds} \right) \frac{dx^\mu}{ds} + \tilde{m}_1 \frac{d^2 x^\mu}{ds^2} + \tilde{m}_1 \Gamma^\nu_{\mu \lambda} \frac{dx^\mu}{ds} \frac{dx^\lambda}{ds}  \\
			&+ 1 \leftrightarrow 2,
		\end{split}
	\end{align}
	where $\tilde{m} _1= m_1 \delta(x - y_1)$,the first term on the right-hand side corresponds to the continuity equation of matter flow $\nabla_\mu (\rho u^\mu) = 0$ in the case of constant mass, but in the case where the mass is time-dependent
	
	For body 1, this gives an effective equation of motion of the form
	\begin{equation}
		\frac{d^2 x^\lambda}{ds^2}
		+\Gamma^\lambda_{\mu\nu}
		\frac{dx^\mu}{ds}
		\frac{dx^\nu}{ds}
		+\frac{1}{m_1}
		\frac{d m_1}{ds}
		\frac{dx^\lambda}{ds}
		=0 .
		\label{Eq:EOM_MT}
	\end{equation}
	
	For the PN expansion, it is useful to rewrite Eq.~\eqref{Eq:EOM_MT} with
	coordinate time \(t\) as the evolution parameter. Projecting with
	\(g_{i\lambda}\), multiplying by \(ds/dt\), and using
	\begin{equation}
		\frac{ds}{dt} = \sqrt{-g_{\rho\sigma}v^\rho v^\sigma},
	\end{equation}
	we obtain
	\begin{align}
		g_{i\lambda} \frac{d^2 x^\lambda}{ds^2} + g_{i\lambda} \Gamma^\lambda_{\mu \nu} \frac{dx^\mu}{ds} \frac{dx^\nu}{ds} + g_{i\lambda} \frac{1}{m_1} \frac{d m_1}{ds} \frac{dx^\lambda}{ds} = 0.
	\end{align}
	Multiply both sides by $ds/dt$ and use $ds/dt = \sqrt{-g_{\rho \sigma} v^\rho v^\sigma}$. Finally, we have
	\begin{align}
		\begin{split}
		&\frac{d}{dt} (g_{i\lambda}  v^\lambda /  \sqrt{- g_{\rho \sigma} v^\rho v^\sigma}) -\frac{1}{2} \partial_i g_{\mu\nu} v^\mu v^\nu /  \sqrt{- g_{\rho \sigma} v^\rho v^\sigma} \\
		& + \frac{\dot{m}_1}{m_1} g_{i\lambda} v^\lambda /\sqrt{- g_{\rho \sigma} v^\rho v^\sigma}  = 0.
		\end{split}
	\end{align}
	Define the Newtonian-like equation of motion as follows:
	\begin{align}  \label{dP/dt=F}
		\frac{d {\mathcal{P}}}{dt} &= {F}_G + {F}_M ,
	\end{align}
	The “linear momentum” vector $\mathcal{P}$ , “gravitational force” (per unit of mass) $\mathcal{F}_G$ and "mass transfer force" ${F}_M$ are defined by
	\begin{align}
		\mathcal{P}^i &= 	g_{i\lambda}  v^\lambda /  \sqrt{- g_{\rho \sigma} v^\rho v^\sigma} ,\\
		{F}_G^i &=  \frac{1}{2} \partial_i g_{\mu\nu} v^\mu v^\nu /  \sqrt{- g_{\rho \sigma} v^\rho v^\sigma} , \\
		{F}_M^i &= - \frac{\dot{m}_1}{m_1} g_{i\lambda} v^\lambda /\sqrt{- g_{\rho \sigma} v^\rho v^\sigma}  .
	\end{align}
	Substituting the 1PN metric into Eq.\eqref{dP/dt=F}, we obtain 
	\begin{align}
		\boldsymbol{a}_1 &= -\frac{G m_2}{r^2} \Bigl[ \bigl(1 + \mathcal{A}_0 + \mathcal{A}_{MT}\bigr) \boldsymbol{n} + \bigl(\mathcal{B}_0 + \mathcal{B}_{MT}\bigr) \boldsymbol{v} \Bigr] \\ \notag
		& + \Bigl( -\frac{\dot{m}_1}{m_1} + \mathcal{C}_{MT} \Bigr) \boldsymbol{v}_1 + \mathcal{O}(c^{-4}),
	\end{align}
	where $\mathbf{n} = \mathbf{x}/r$, orbital separation and relative position $r = r_{12} = |x_1 - x_2|, r' = v \cdot n $ and relative velocity
	$v = v_A - v_B = dx/dt$. The coefficients are given by
	\begin{align}
		\begin{split}
		\mathcal{A}_0 &= \frac{1}{c^2} \biggl[-5 \frac{G m_1}{r} - 4 \frac{G m_2}{r} - \frac{3}{2} \bigl(\boldsymbol{v}_2 \cdot \boldsymbol{n}\bigr)^2   \\
		& \qquad  \quad- 4 \boldsymbol{v}_1 \cdot \boldsymbol{v}_2 + v_1^2 + 2 v_2^2 \biggl]  , \end{split}\\
		\mathcal{A}_{MT} &= \frac{1}{c^2} \left[ -\frac{3}{4} r \bigl(\boldsymbol{v}_2 \cdot \boldsymbol{n}\bigr) \frac{\dot{m}_1}{m_1} + \frac{1}{4} r \bigl(\boldsymbol{v}_2 \cdot \boldsymbol{n}\bigr) \frac{\dot{m}_2}{m_2} \right],\\
		\mathcal{B}_0 &= \frac{1}{c^2} \biggl[ -4 (\boldsymbol{v}_1 \cdot \boldsymbol{n}) + 3 (\boldsymbol{v}_2 \cdot \boldsymbol{n}) \biggl],\\
		\mathcal{B}_{MT} &= \frac{1}{c^2} \left[ \frac{13}{4} r \frac{\dot{m}_1}{m_1} + \frac{1}{4} r \frac{\dot{m}_2}{m_2} \right], \\
		\mathcal{C}_{MT} &= \frac{1}{c^2} \left[ v_1^2 \frac{\dot{m}_1}{m_1} + \frac{G m_2}{r} \left( \frac{13}{4} \frac{\dot{m}_1}{m_1} - \frac{11}{4} \frac{\dot{m}_2}{m_2} \right) \right].
	\end{align}
	We now derive the 1PN order equations of motion in the center-of-mass frame. Transforming the relative acceleration $a = a_A - a_B$into the center-of-mass frame, we have
	\begin{widetext}
	\begin{align}
		\boldsymbol{a} &= -\frac{G m}{r^2} \Bigl[ \bigl(1 + \tilde{\mathcal{A}}_0 + \tilde{\mathcal{A}}_{MT} \frac{\dot{\mu}}{\mu} \bigr) \boldsymbol{n} + \bigl(\tilde{\mathcal{B}}_0 + \tilde{\mathcal{B}}_{MT} \frac{\dot{\mu}}{\mu} \bigr) \boldsymbol{v} \Bigr] + \frac{\dot{\mu}}{\mu} \bigl(1 + \mathcal{C}_{MT}\bigr) \boldsymbol{v} + \mathcal{O}(c^{-4})
	\end{align}
	\begin{align}
		\begin{split}
		\tilde{\mathcal{A}}_0 &= \frac{1}{c^2} \left[ -\frac{3}{2} \dot{r}^2 \eta + v^2 + 3\eta v^2 - \frac{G M}{r} \bigl(4 + 2\eta\bigr) \right] ,\qquad \tilde{\mathcal{A}}_{MT} = \frac{1}{c^2} \biggl[ -\eta r \dot{r} \biggl] ,\\
		\tilde{\mathcal{B}}_0 &= \frac{1}{c^2} \biggl[  -4 \dot{r} + 2 \dot{r} \eta \biggl]  ,\qquad \tilde{\mathcal{B}}_{MT} = \frac{1}{c^2} \biggl[ -r  \biggl] ,\qquad  \tilde{\mathcal{C}}_{MT} = \frac{1}{c^2} \biggl[  -v^2 + \frac{3}{2} v^2 \eta \biggl] .
		\end{split}
	\end{align}
	\end{widetext}
	where $\mu = m_1 m_2/m$ is the reduced mass , $\eta = \mu / m$ is the symmetry mass ratio. This result is consistent with that obtained in Ref.~\cite{Zhang20252}.
	\subsection{Kepler's Third Law}
	We restrict the discussion to circular binaries and use the angular-momentum
	balance equation. The relative velocity can be decomposed as  $\boldsymbol{v}  = \dot{r} \boldsymbol{n}+  \omega r \mathbf{\tau} $, $\mathbf{\tau}$ is the unit tangent vector in the orbit plane and $\omega$ is the angular velocity of binary.
	We can get the motion in this frame as
	\begin{align}
		\dot{n} = \frac{1}{r} (\boldsymbol{v}  - \dot{r} \boldsymbol{n} ) = \omega \boldsymbol{\tau}, \qquad \dot{\boldsymbol{\tau}} = - \omega \boldmath{n},
	\end{align}
	for the $\boldsymbol{n}$ direction acceleration we get
	\begin{align}
		\omega^2 r  =  -\frac{G m}{r^2} \bigl(1 + \tilde{\mathcal{A}}_0 + \tilde{\mathcal{A}}_{MT} \frac{\dot{\mu}}{\mu} \bigr) .
	\end{align}
	We get the Kepler third law with MT correction at the 1PN order as

	\begin{align}
		\omega^2 = - \frac{Gm}{r^3}  \biggl[ 1 + \frac{1}{c^2} \biggl(& -\frac{3}{2} \dot{r}^2 \eta + v^2 + 3\eta v^2  \notag\\
		& - \frac{G m}{r} \bigl(4 + 2\eta\bigr) -\eta r \dot{r} \frac{\dot{\mu}}{\mu}  \biggl) \biggl].
	\end{align}
	\section{Gravitational Radiation with Mass Transfer}
	
	In this section, we compute the gravitational-wave energy and angular-momentum
	fluxes for a binary undergoing mass transfer. We use the standard multipolar
	post-Minkowskian expansion for the radiation field \cite{Thorne1980,Will1996,Blanchet2024}. The standard compact-binary inspiral driven by gravitational radiation is described by Peters and Mathews \cite{Peters1963,Peters1964}. The source multipole moments are evaluated for a binary whose component masses vary slowly in time.
	
	We work in the adiabatic mass-transfer regime. The mass-transfer timescale is
	assumed to be much longer than the orbital timescale. We therefore retain terms
	proportional to the first time derivatives of the mass parameters, such as
	\(\dot\mu\), and neglect higher derivatives such as \(\ddot\mu\).
	
	Mass transfer affects the radiation sector in two ways. First, the orbital
	dynamics is modified by the changing masses. This changes \(r(t)\),
	\(\omega(t)\), and the orbital phase. Second, the source multipole moments
	depend explicitly on the time-dependent masses. Their time derivatives then
	generate additional terms proportional to \(\dot\mu\) and related mass
	derivatives.
	
	The balance equations can be written schematically as
	\begin{align}
		\frac{d \mathcal{H}}{dt} &= -\mathcal{F}_{\rm GW}, \\
		\frac{d \mathcal{J}^i}{dt} &= -\mathcal{G}^i_{\rm GW}.
	\end{align}
	Here \(\mathcal{H}\) and \(\mathcal{J}^i\) denote the orbital energy and angular momentum, respectively. 
	
	Using the leading multipolar post-Minkowskian relation between radiative and
	source multipoles, and keeping the source multipoles to the order required for
	1PN accuracy, we obtain
	\begin{align}
		\mathcal{F}_{GW} =& \frac{G}{5c^5} \big\langle \hat I^{(3)}_{ij}\hat I^{(3)}_{ij} \big\rangle \notag\\
		&+ \frac{G}{c^7} \left[ \frac{16}{45} \big\langle \hat J^{(3)}_{ij}\hat J^{(3)}_{ij}
		\big\rangle + \frac{1}{189} \big\langle
		\hat I^{(4)}_{ijk}\hat I^{(4)}_{ijk} \big\rangle \right] +\mathcal{O}(c^{-9}), \\
		\mathcal{G}_{GW}^i =& \frac{2G}{5c^5} \epsilon_{ijk} \big\langle \hat I^{(2)}_{jl}\hat I^{(3)}_{kl}
		\big\rangle \notag\\
		&+ \frac{G}{c^7} \epsilon_{ijk} \left[ \frac{32}{45} \big\langle
		\hat J^{(2)}_{jl}\hat J^{(3)}_{kl} \big\rangle
		+ \frac{1}{63} \big\langle \hat I^{(3)}_{jlm}\hat I^{(4)}_{klm} \big\rangle \right] +\mathcal{O}(c^{-9}) .
	\end{align}
	The derivation and conventions for the multipole expansion are summarized in
	Appendix~\ref{Multipole}.
	
	In the center-of-mass frame, the mass quadrupole at 1PN order is
	\begin{align}
		\hat I_{ij} &= \mu x_{\langle i}x_{j\rangle} \notag\\
		&\quad +\frac{\mu}{c^2}
		\biggl\{ \left[	\frac{29}{42}(1-3\eta)v^2 -
		\frac{1}{7}(5-8\eta)\frac{Gm}{r}	\right]x_{\langle i}x_{j\rangle} \notag\\
		&\qquad- \frac{4}{7}(1-3\eta)r\dot r\,x_{\langle i}v_{j\rangle}
		+ \frac{11}{21}(1-3\eta)r^2v_{\langle i}v_{j\rangle}	\biggr\}.
	\end{align}
	Angular brackets denote the symmetric trace-free projection. For example,
	\begin{equation}
		x_{\langle i}x_{j\rangle} = x_{(i}x_{j)} - \frac{1}{3}\delta_{ij}x_kx^k .
	\end{equation}
	The current quadrupole and mass octupole are needed only at Newtonian order:
	\begin{align}
		\hat J^{ij}
		&=
		\mu \sqrt{1-4\eta}\,
		\epsilon^{kl\langle i}x^{j\rangle}x^k v^l, \\
		\hat I^{ijk}
		&=
		\mu \sqrt{1-4\eta}\,
		x^{\langle i}x^j x^{k\rangle}.
	\end{align}
	Substituting these moments into the flux formulae gives
	\begin{widetext}
	\begin{align}
		\begin{split}
			\mathcal{F}_{GW} &= \frac{1}{c^5} \frac{32}{5}G \mu^2 (G m) ^{4/3} \omega^{4/3} \biggl\{ \omega^2 + \frac{1}{3} \Big( \frac{\dot{\mu}}{\mu} \Big)^2+ \frac{1}{c^2} (Gm \omega)^{2/3} \biggl[ - \Big(\frac{1247}{336} + \frac{35}{12} \eta\Big) \omega^{2} 
			-\Big(\frac{577}{504} + \frac{11}{16}\eta \Big)  \Big( \frac{\dot{\mu}}{\mu} \Big)^2 
			\biggl]\biggl\},
		\end{split} \\
		\mathcal{G}^i_{GW} &= \frac{1}{c^5} \frac{32}{5}G \mu \mathcal{J}^i (G m )^{2/3} \omega^{2/3} \biggl\{ \omega^2 + \frac{1}{4} \Big( \frac{\dot{\mu
		}}{\mu}\Big)^2 + \frac{1}{c^2} (Gm \omega )^{2/3} \biggl[ -(\frac{575}{336} + \frac{43}{12}\eta ) \omega^{2} - \Big(  \frac{109}{504} + \frac{13}{12} \eta \Big) \Big( \frac{\dot{\mu}}{\mu} \Big)^2 
		\biggl] \biggl\} .
	\end{align}
	\end{widetext}
	\section{Waveform with Mass Transfer}
	
	In this section, we construct the far-zone gravitational waveform from a
	mass-transferring binary. The waveform is kept to 1PN order. The effects of
	mass transfer enter through the slow time dependence of the masses and through
	the modified orbital evolution.We retain their first time derivatives in the waveform and neglect higher derivatives, such as
	\(\ddot{\mu}\) and \(\ddot{\eta}\).
	
	We define
	\begin{align}
	\mu=\frac{m_1m_2}{m}, \quad \eta=\frac{\mu}{m},\quad \chi=\frac{m_1-m_2}{m}.
	\end{align}
	The quantities \(\mu\), \(\eta\), \(\chi\), the orbital frequency \(\Omega\),
	and the orbital phase \(\psi\) are slowly varying functions of time.
	
	In the far zone, the metric perturbation can be written in terms of the mass
	multipoles \(I_{ij}\), \(I_{ijk}\), and the current multipole \(J_{ij}\) as
	\begin{align}
		h_{jk} = \frac{2G}{D c^4}
		\biggl[ \ddot I_{jk}
		+\frac{1}{3c} \left( \dddot I_{jkn} +2\epsilon^{mn}{}_{j}\ddot J_{mk} +2\epsilon^{mn}{}_{k}\ddot J_{mj} \right)N_n \biggr] .
	\end{align}
	Here \(D\) is the distance to the source, and \(N^i\) is the unit vector from
	the source to the observer. The multipoles are evaluated at the retarded time.
	
	The transverse-traceless waveform is obtained by applying the TT projector,
	\begin{equation}
		h^{\rm TT}_{jk} = \Lambda_{jk}{}^{mn}h_{mn},
	\end{equation}
	where
	\begin{equation}
		\Lambda_{jk}{}^{mn}
		= P_j{}^m P_k{}^n -
		\frac{1}{2}P_{jk}P^{mn}, \quad
		P_j{}^k=\delta_j{}^k-N_jN^k .
	\end{equation}
	
	The plus and cross polarizations can be written as
	\begin{align}
		&h_{+,\times} = \frac{2 G \mu}{c^2 D} \Big( \frac{G m \omega}{c^3} \Big) ^{2/3} H_{+,\times}, \\
		H_{+,\times} &= \underset{0 \quad \;}{H_{+,\times }} +\chi \beta \underset{1/2 \quad \;}{H_{+,\times}}+ \beta^2\underset{1 \quad \;}{H_{+,\times}} + \mathcal{O}_{\beta^3},
	\end{align}
	where $ \beta = ( {Gm\omega} / {c^3})^{1/3}. $
	
	Let \(c_i=\cos i\) and \(s_i=\sin i\), where \(i\) is the inclination angle
	between the orbital angular momentum and the line of sight. $\psi$ is the orbit phase. The leading
	polarization modes, including the explicit mass-transfer terms from
	\(\dot\eta\) and \(\dot\chi\), are then obtained by differentiating the
	time-dependent source multipoles. 
	\begin{widetext}
	For the plus and cross polarization we have
	\begin{align}
		\underset{0 \; \;}{H_+} &= -(1 + c_i^2)\cos 2\psi -\frac{1}{2} \frac{\dot{\eta}}{\eta \omega} (1 + c_i^2) \sin 2\psi ,   
		 \qquad \qquad    &	\underset{0 \; \;}{H_\times} &= -2 c_i \sin 2\psi  + c_i \cos 2\psi \frac{\dot{\eta}}{\eta \omega},\\
		\underset{1/2 \;}{H_+} &= -s_i\chi\Bigl[\cos\psi\bigl(\frac{5}{8} + \frac{1}{8}c_i^2\bigr) - \cos 3\psi\bigl(\frac{9}{8} + \frac{9}{8}c_i^2\bigr)\Bigr]  
		& 	\underset{1/2 \;}{H_\times} &= s_i c_i \chi \biggl[ -\frac{3}{4} \sin \psi + \frac{9}{4} \sin 3\psi \biggl],  \\ \notag
		& \quad + s_i\chi \left[\sin\psi( \frac{7}{2} - \frac{5 }{2}c_i^2)  - \sin3\psi (\frac{1}{2} + \frac{c_i^2}{2})  \right] \frac{\dot{\eta}}{\eta \omega}  
		&& \quad + s_i c_i\chi \left[- \frac{1}{4} \cos \psi  +\frac{1}{4} \cos 3\psi   \right] \frac{\dot{\eta}}{\eta \omega}  \\ \notag
		& \quad + s_i  \dot{\chi} \left[ \sin\psi (-5 - c_i^2) + \sin3\psi (3 + 3c_i^2) \right] \frac{1}{\omega}, && \quad + s_i  c_i\dot{\chi} \left[ - \frac{1}{4}\cos \psi + \frac{1}{4} \cos 3\psi \right] \frac{1}{\omega} ,\\ \notag
		\underset{1 \;\;}{H_+} &= \cos 2\psi\Bigl[\frac{19}{6} + \frac{3}{2}c_i^2 - \frac{1}{3}c_i^4 + \eta\Bigl(-\frac{19}{6} + \frac{11}{6}c_i^2 + c_i^4\Bigr)\Bigr]  &\underset{1 \;\;}{H_\times} &= c_i \sin 2\psi\Bigl[\frac{17}{3} - \frac{4}{3} c_i^2 + \eta\Bigl(-\frac{13}{3} + 4 c_i^2\Bigr)\Bigr] .\notag\\ 
		 & - \cos 4\psi\Bigl[\frac{4}{3}s_i^2(1 + c_i^2)(1 - 3\eta)\Bigr] .
	\end{align}
	\end{widetext}
	
	Mass transfer enters the waveform in two ways. First, the amplitude depends on
	the time-dependent reduced mass. Second, the phase evolution is
	modified by the corrected orbital frequency and radiation-reaction balance.
	Additional terms proportional to \(\dot{\chi}\) and \(\dot{\eta}\) arise when
	time derivatives act on the source multipole moments. These terms appear
	explicitly in the polarization modes.
	
	For the systems considered here, the instantaneous waveform corrections are
	small. The main effect is instead a secular phase shift accumulated over many
	orbital cycles. This effect is relevant for ultracompact binaries in the
	millihertz band, where space-based detectors may observe the source over
	multi-year timescales \cite{Cutler1998,Cornish2003,AmaroSeoane2023,Robson2019,Kupfer2018}. Similar secular signatures have been emphasized in recent mass-transfer waveform and detectability studies \cite{Zhang2024,Zhang20251,Zhang20262}.
	
	\section{Numerical results \label{NumR}}
	
	We use the numerical examples for three purposes. First, we show that the
	long-term Roche-lobe-overflow evolution is self-regulated after contact.
	Second, we compare the mass-transfer contribution to the orbital-period
	derivative with the gravitational-wave contribution. Third, we estimate the
	time required for mass transfer to produce an observable waveform phase shift.
	
		Upon reaching the onset of Roche lobe overflow , the system can maintain stable mass transfer over a long timescale. In practice, the RLO model can be categorized into optically thin and optically thick regimes. We will simulate the instantaneous mass transfer rates between the binary components under both of these conditions.
	    
	     	We consider two distinct scenarios: 
	     	
	     	(i) The white dwarf radius fully fills the Roche lobe, simulating a more advanced stage of mass transfer. 
	     	
	     	(ii) The white dwarf radius marginally exceeds the Roche lobe radius, representing the initial onset of mass transfer.
	     	
	  \begin{figure}[h]
	     \centering
	     \includegraphics[width=\linewidth]{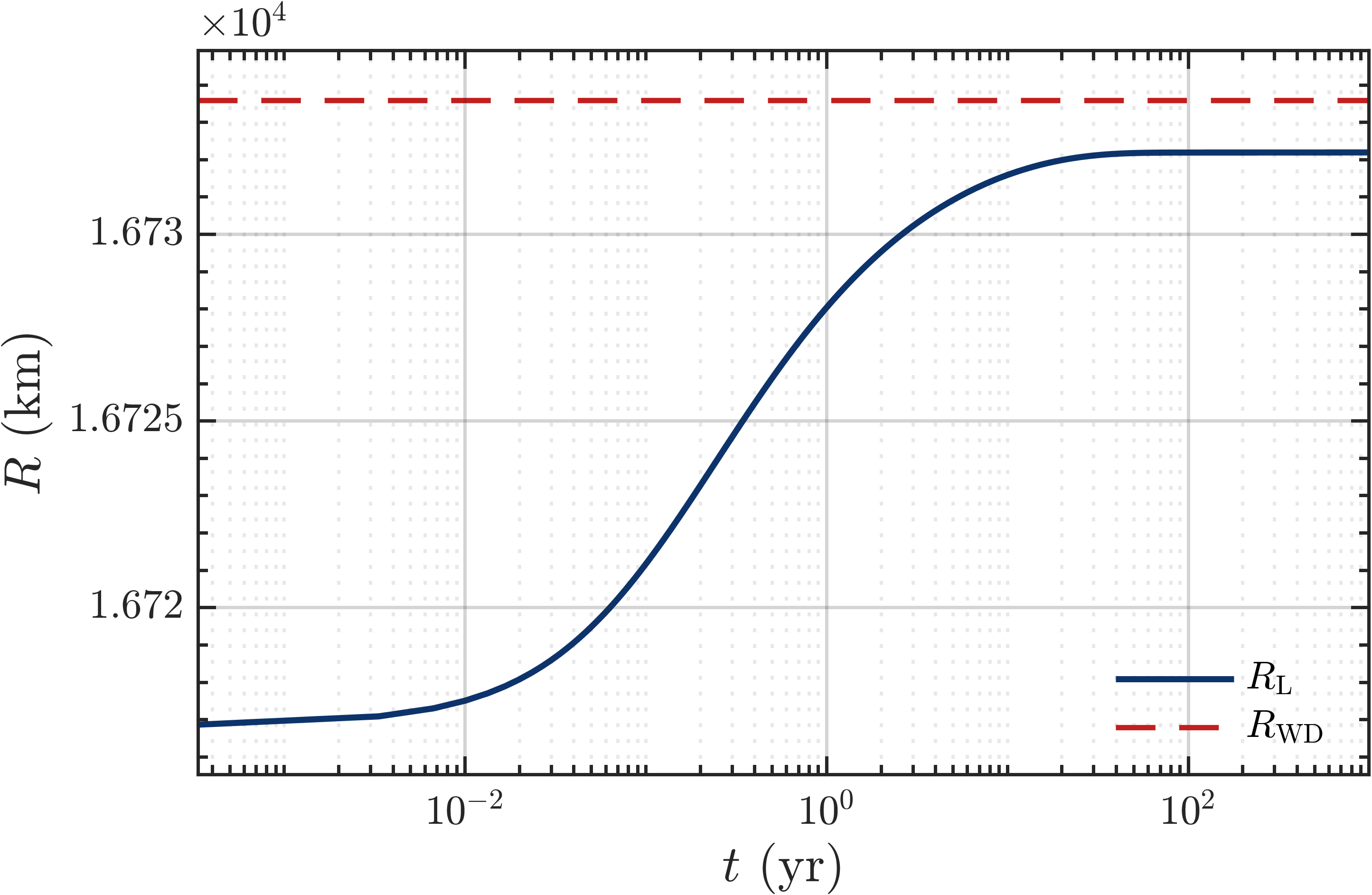} \\
	     \includegraphics[width=\linewidth]{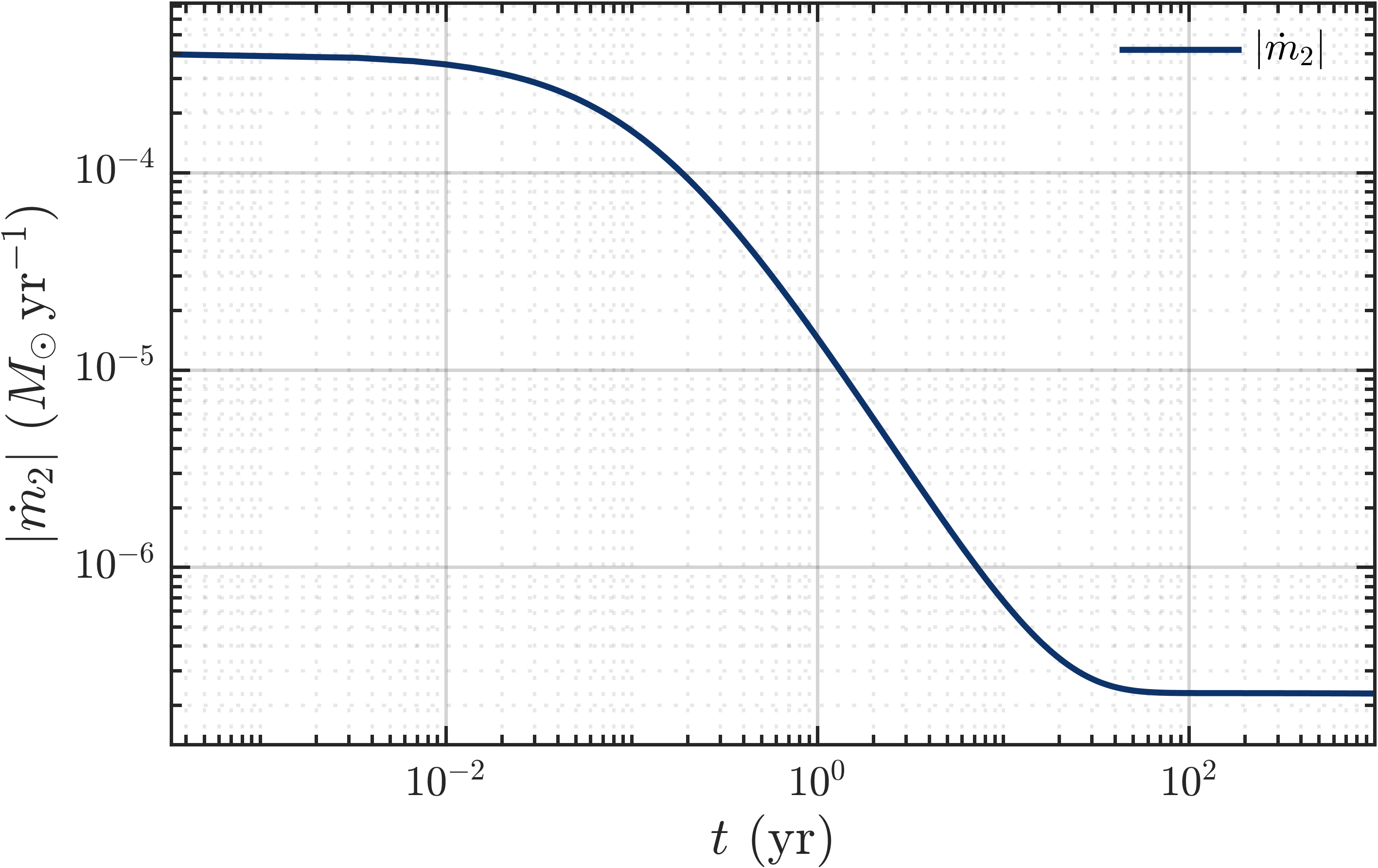}
	    \caption{Long-term evolution of an initially overfilled binary. The panels show the Roche-lobe radius relative
	    	to the white-dwarf radius and the corresponding mass-transfer rate. After the initial transient, the binary relaxes toward marginal contact and a lower quasi-steady mass-transfer rate.}
	     \label{fig:case1}
	  \end{figure}

	Figure~\ref{fig:case1} shows the initially overfilled case.
	In the initially overfilled case, the mass-transfer rate is large at early times. The orbit expands because mass is transferred from the lighter donor to the heavier accretor. The Roche-lobe radius then moves toward the white-dwarf
	radius, and \(|\dot M_2|\) rapidly decreases to a lower quasi-steady value.
	
		\begin{figure}[h]
		\centering
		\includegraphics[width=\linewidth]{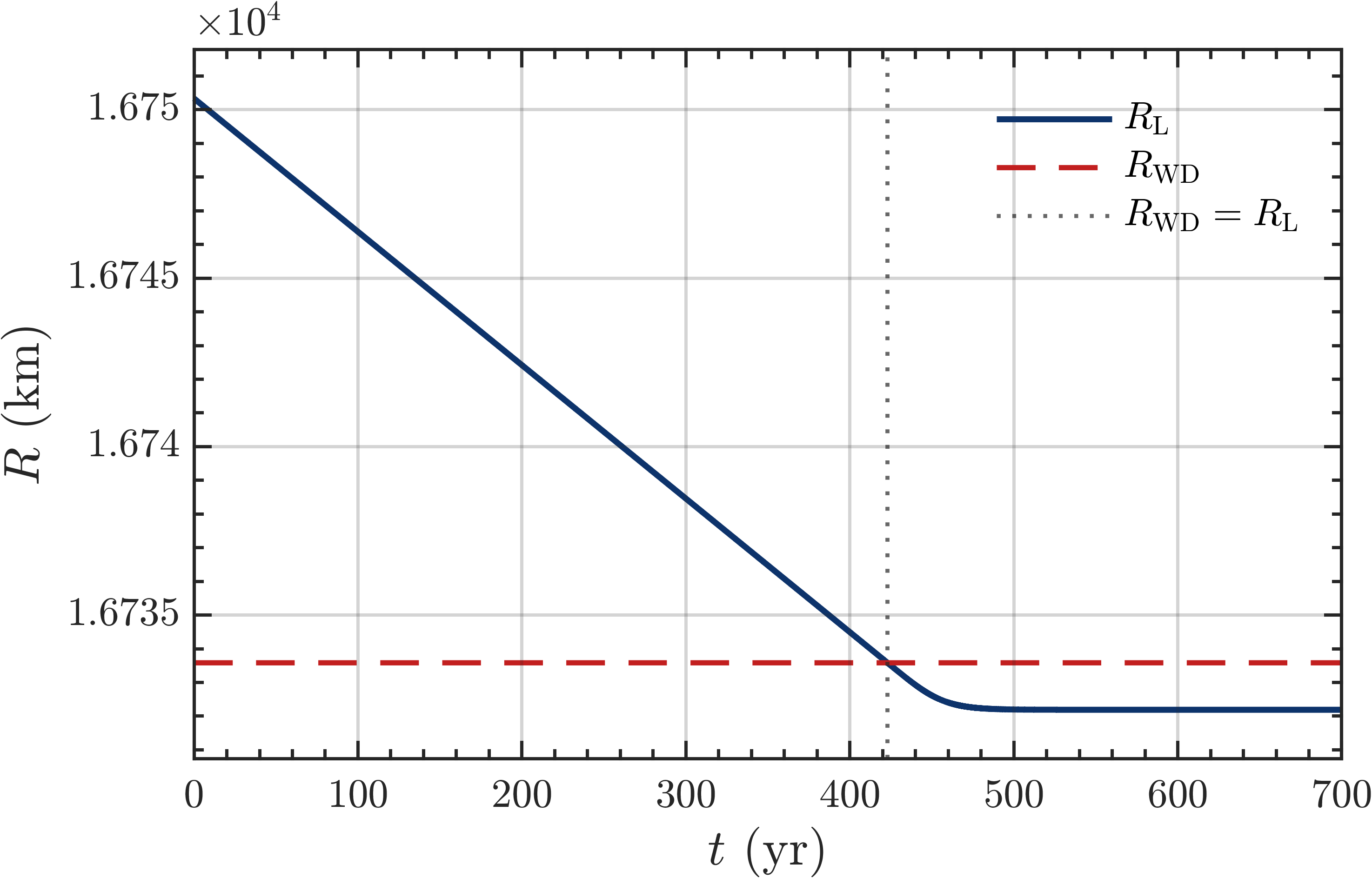} \\
		\includegraphics[width=\linewidth]{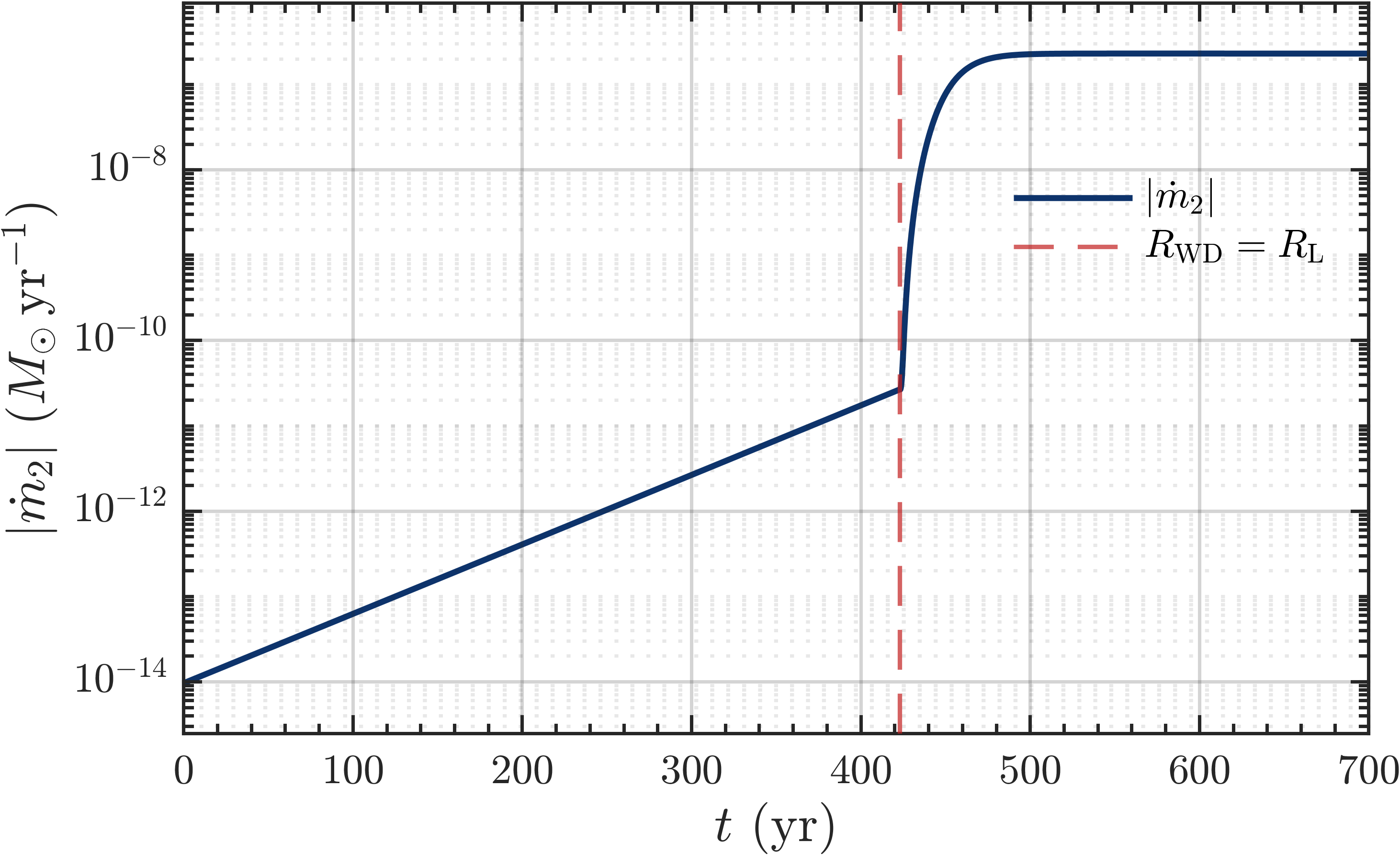}
		\caption{Long-term evolution of an initially underfilled binary. The panels show the shrinking Roche-lobe
			radius before contact and the subsequent rise of the mass-transfer rate after the donor reaches Roche-lobe overflow.}
		\label{fig:case2}
	\end{figure}     	
	The initially underfilled case evolves in a different way. At early times the
	binary is almost detached, and gravitational-wave emission removes orbital
	angular momentum \cite{Peters1963,Peters1964}. The Roche lobe therefore shrinks until the donor reaches
	contact. After contact, the mass-transfer rate rises rapidly and approaches the
	same regulated branch found in the initially overfilled calculation. This
	comparison shows that the later RLO evolution is not determined only by the
	initial fill-out factor. It is controlled by the competition between
	gravitational-wave-driven orbital shrinkage and mass-transfer-driven orbital
	expansion.
	
	\begin{table*}[htb]
		\centering
		\caption{Period derivative contributions for the slightly overfilled mass-transferring binary. The initial condition is $R_{\rm WD}/R_{\rm L}-1=10^{-3}$. Ratios are normalized by $|\dot{P}_{\rm N\,GW}|$ at the same epoch.}
		\label{tab:overfill-pdot-contributions}
		\begin{tabular}{p{3cm}|p{3cm}|p{2.5cm}|p{3cm}|p{2.5cm}}
			\hline\hline
			\multicolumn{1}{c|}{the MT rate $|\dot{m}_2|$} & \multicolumn{2}{c|}{$3.98\times 10^{-4}\,M_\odot\,{\rm yr}^{-1}$} & \multicolumn{2}{c}{$6.75\times 10^{-7}\,M_\odot\,{\rm yr}^{-1}$} \\
			\hline
			\multicolumn{1}{c|}{$( R_{\rm WD} - R_{\rm L})/R_{\rm L}$} & \multicolumn{2}{c|}{$1\times 10^{-3}$} & \multicolumn{2}{c}{$1.19\times 10^{-4}$} \\
			\hline
			Term & \multicolumn{2}{c|}{Initial} & \multicolumn{2}{c}{After 10 yr} \\
			\hline
			& $\dot{P}$ (s\,s$^{-1}$) & Rel. to N GW & $\dot{P}$ (s\,s$^{-1}$) & Rel. to N GW \\
			\hline\hline
			Newtonian GW & $-3.63\times 10^{-11}$ & $1$ & $-3.61\times 10^{-11}$ & $1$ \\
			1PN GW  & $2.24\times 10^{-15}$ & $6.18\times 10^{-5}$ & $2.23\times 10^{-15}$ & $6.17\times 10^{-5}$ \\
			MT & $7.6\times 10^{-8}$ & $2.1\times 10^{3}$ & $1.29\times 10^{-10}$ & $3.58$ \\
			GW+MT & $7.6\times 10^{-8}$ & $2.1\times 10^{3}$ & $9.33\times 10^{-11}$ & $2.58$ \\
			\hline
		\end{tabular}
	\end{table*}
	
	We next apply the same prescription to 4U 1728-34. The nature of this source is
	still debated, and we do not assume that it is already established as an
	ultracompact X-ray binary \cite{Galloway2010,Vincentelli2020,Vincentelli2023}. Instead, we use it as a consistency test. If the
	observed accretion luminosity is powered mainly by accretion onto the neutron
	star, the mass-transfer rate can be estimated as
	\begin{equation}
		|\dot m_2| \simeq \frac{L_{\rm acc}R_1}{Gm_1},
	\end{equation}
	up to uncertainties in distance, bolometric correction, anisotropy, and
	radiative efficiency.
	
	\begin{figure}[h]
		\includegraphics[width=\linewidth]{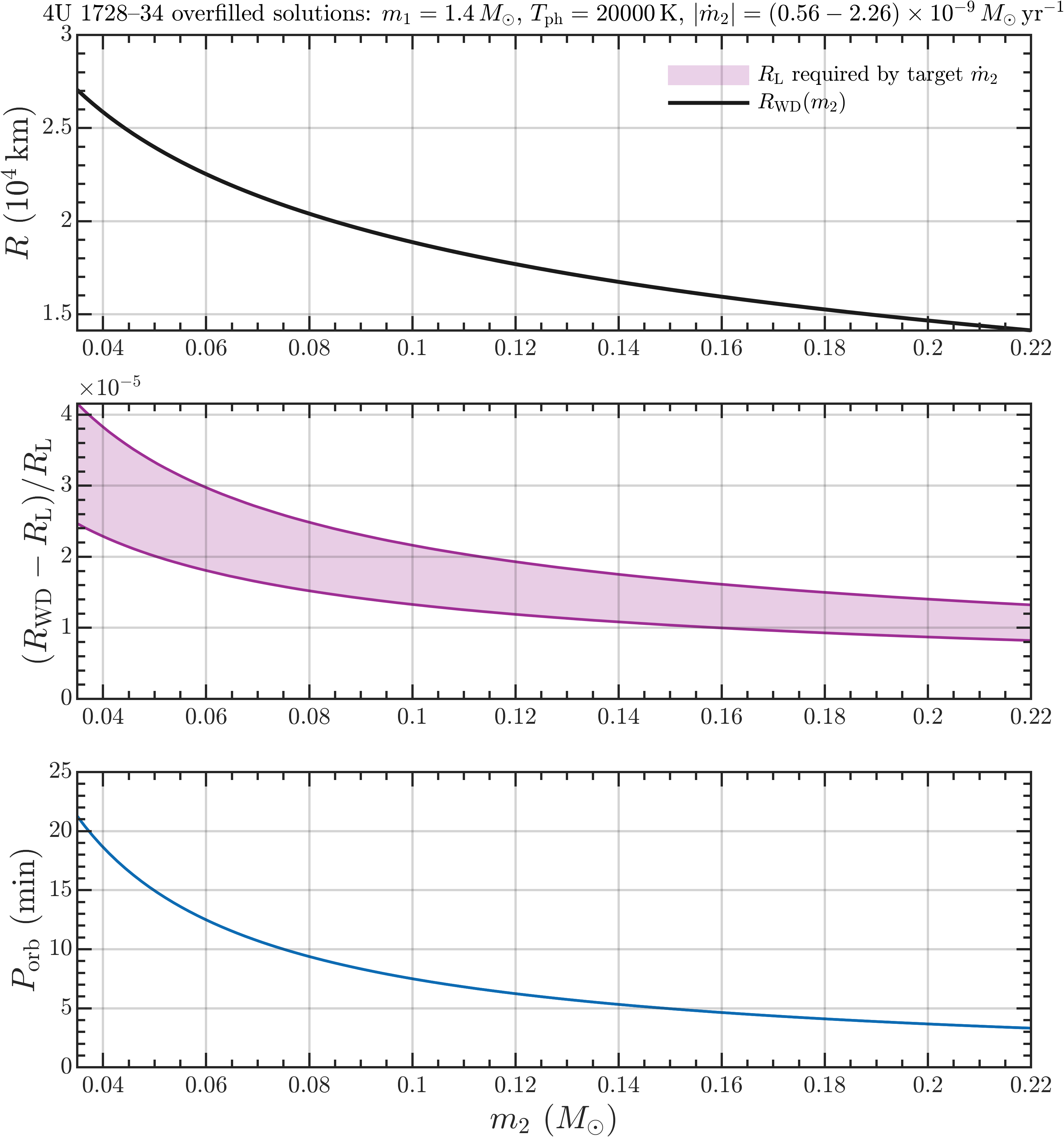}
		\caption{Application of the mass-transfer model to a 4U 1728-34
			system. The target mass-transfer rate is estimated from the accretion
			luminosity. The curves show the white-dwarf radius, the Roche-lobe radius
			required by the target \(|\dot M_2|\), the corresponding overfill factor,
			and the orbital period.}
		\label{fig:4u1728}
	\end{figure}
	
	Figure~\ref{fig:4u1728} shows that, for a neutron-star--white-dwarf
	configuration, the luminosity-inferred range of \(|\dot M_2|\) can be produced
	with a small Roche-lobe overfill \cite{Misanovic2010,Salgundi2026}. This result does not prove that 4U 1728-34 is
	a UCXB. It shows only that, from the mass-transfer requirement alone, such an
	interpretation is not excluded.
	
	Table~\ref{tab:overfill-pdot-contributions} gives the separate contributions
	to the orbital-period derivative for two mass-transfer rates. The mass-transfer
	term is positive for conservative transfer from the lighter donor to the
	heavier accretor. The main uncertainty in \(\dot P\) therefore comes from the mass-transfer prescription.
	
	\begin{figure*}[t]
		\centering
		\includegraphics[width=0.48\textwidth]{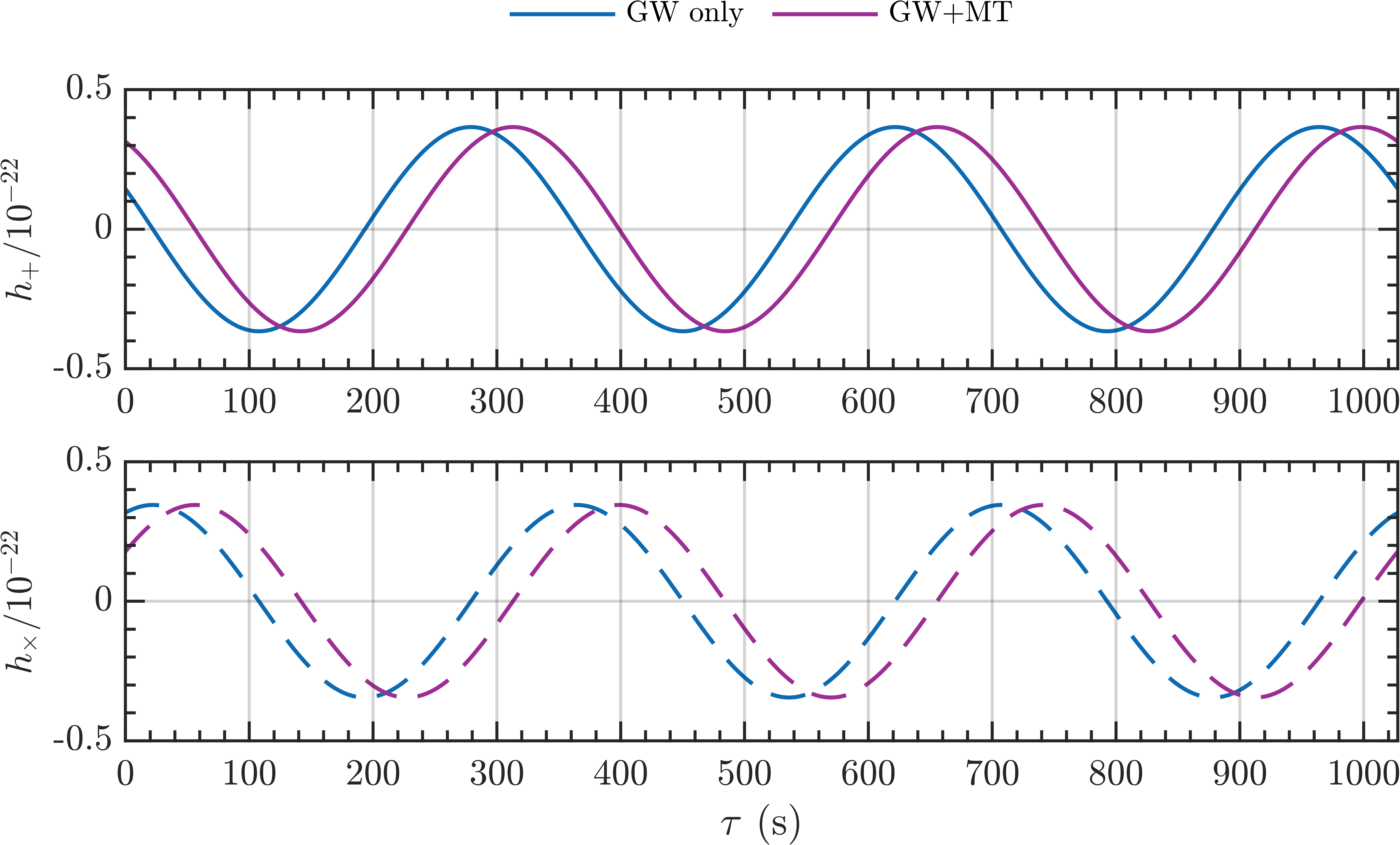}
		\includegraphics[width=0.48\textwidth]{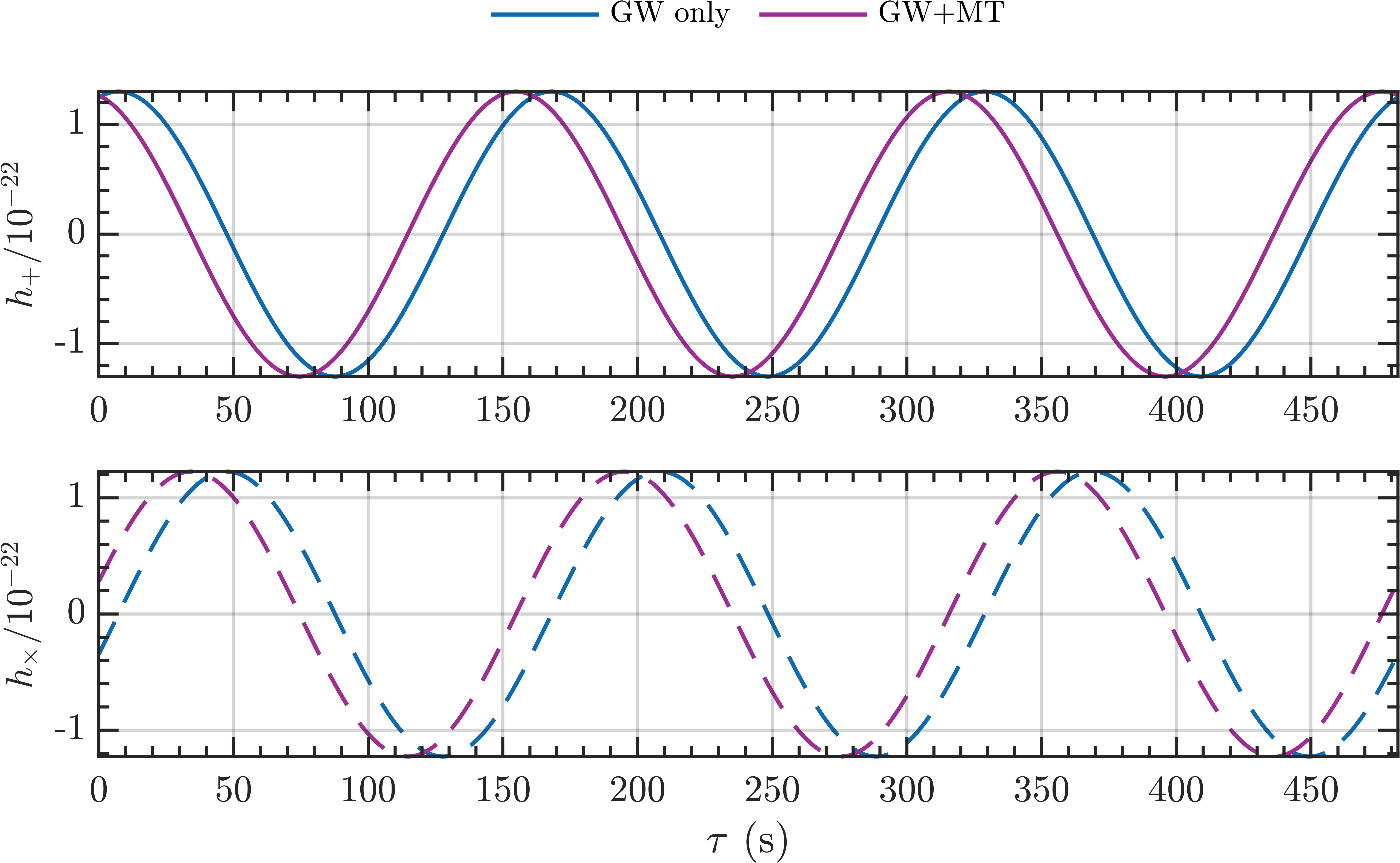}
		\includegraphics[width=0.48\textwidth]{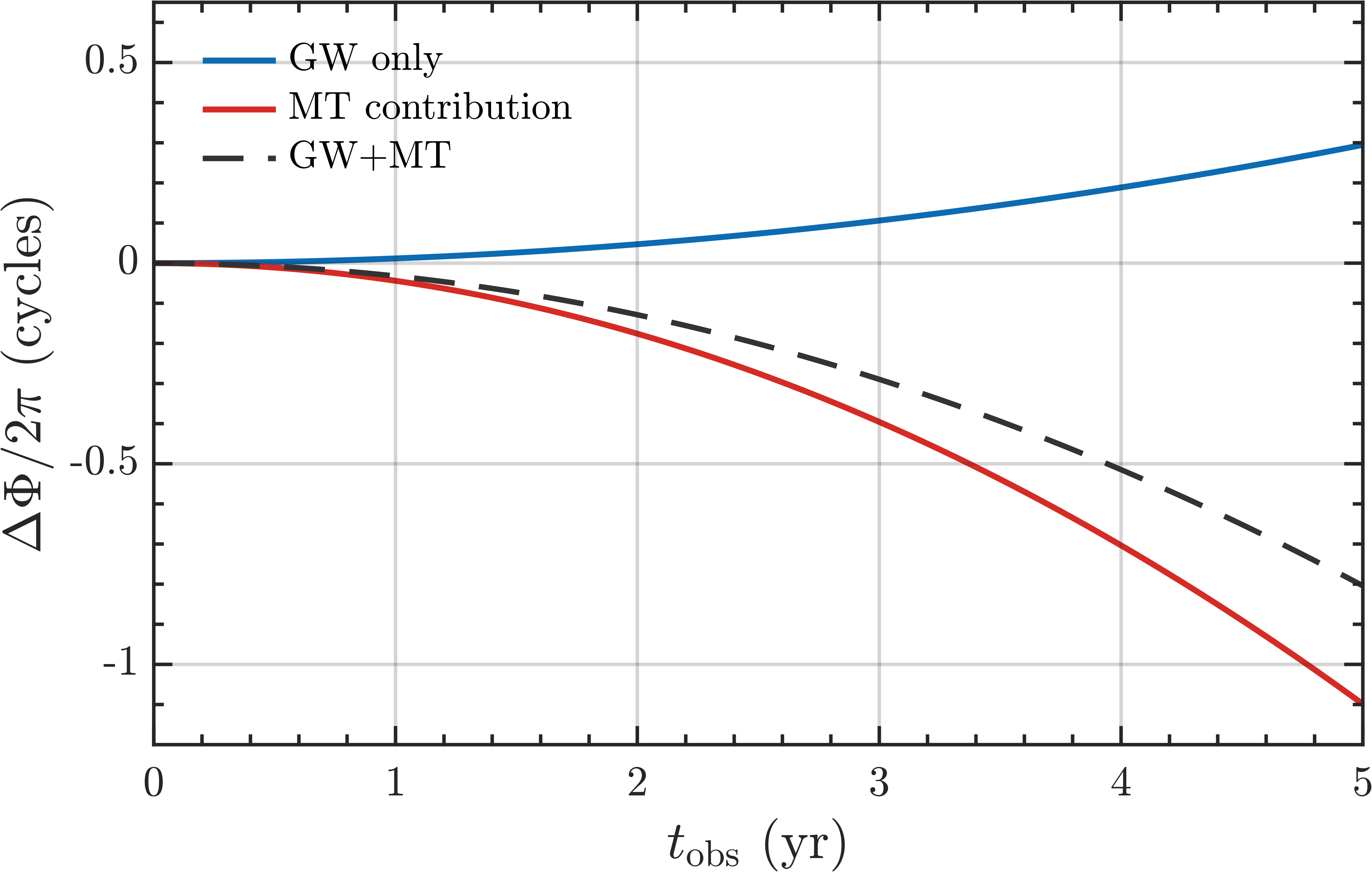}
		\includegraphics[width=0.48\textwidth]{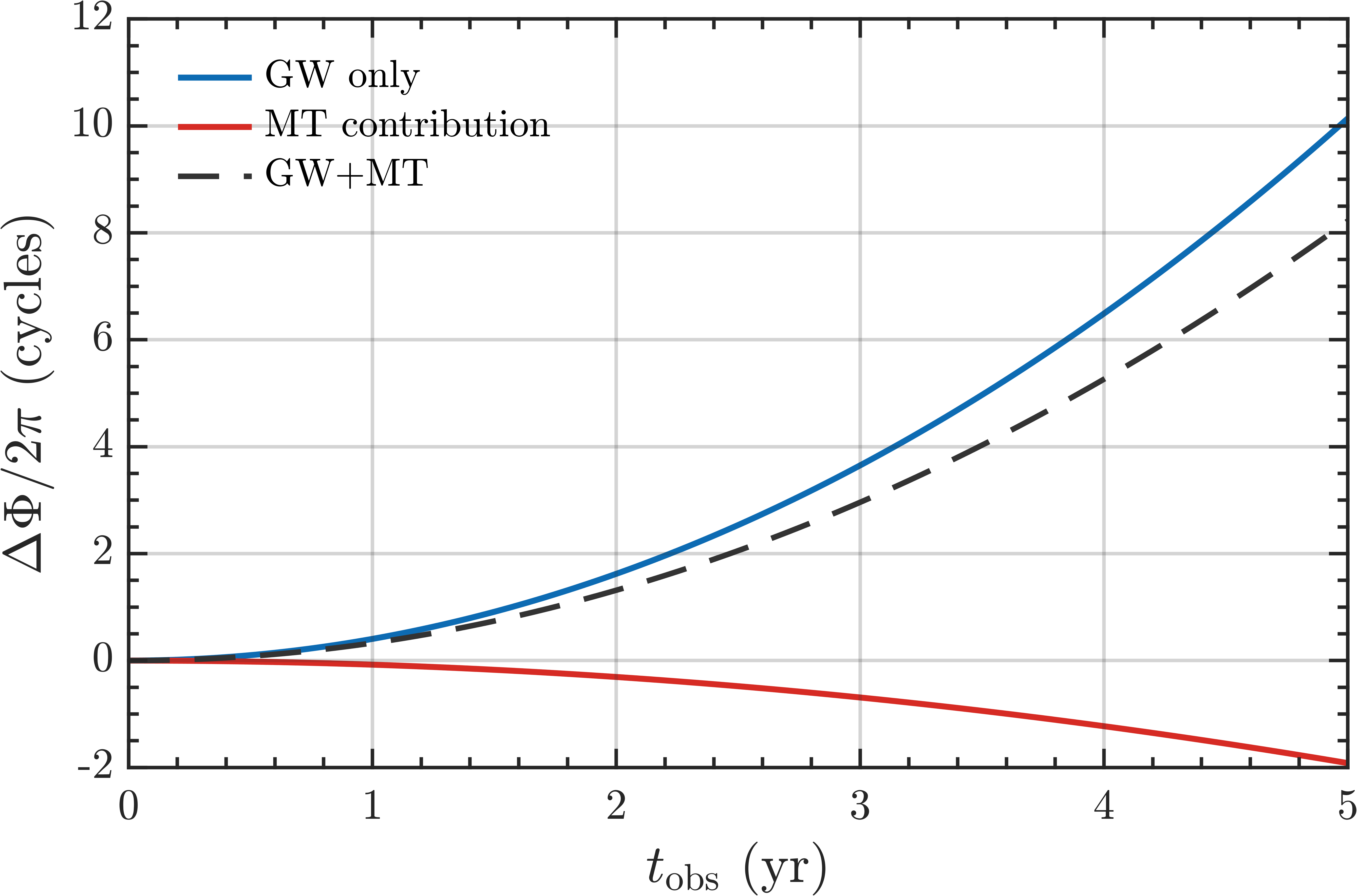}
		\caption{Effect of mass transfer on the gravitational-wave phase for
			4U 1820-30-like and RX J0806.3+1527-like systems. The observed ultrashort
			periods and period derivatives of these sources motivate the adopted
			examples \cite{Stella1987,Chou2001,Peuten2014,Ramsay2002,Strohmayer2005}. The upper panels compare
			local waveforms after the fifth observing year. The lower panels show the
			accumulated GW and MT phase contributions over a five-year observation.}
		\label{fig:waveform-phase}
	\end{figure*}
	
	Figure~\ref{fig:waveform-phase} shows that mass transfer mainly affects the
	waveform through a secular phase shift. The instantaneous amplitudes remain
	similar in the examples considered here. For a 4U 1820-30-like system, the
	mass-transfer contribution can dominate the accumulated phase evolution. For
	an RX J0806.3+1527-like system, gravitational radiation remains the leading
	driver, but mass transfer still partially cancels the inspiral phase.
	\begin{figure*}[t]
		\centering
		\includegraphics[width=0.48\textwidth]{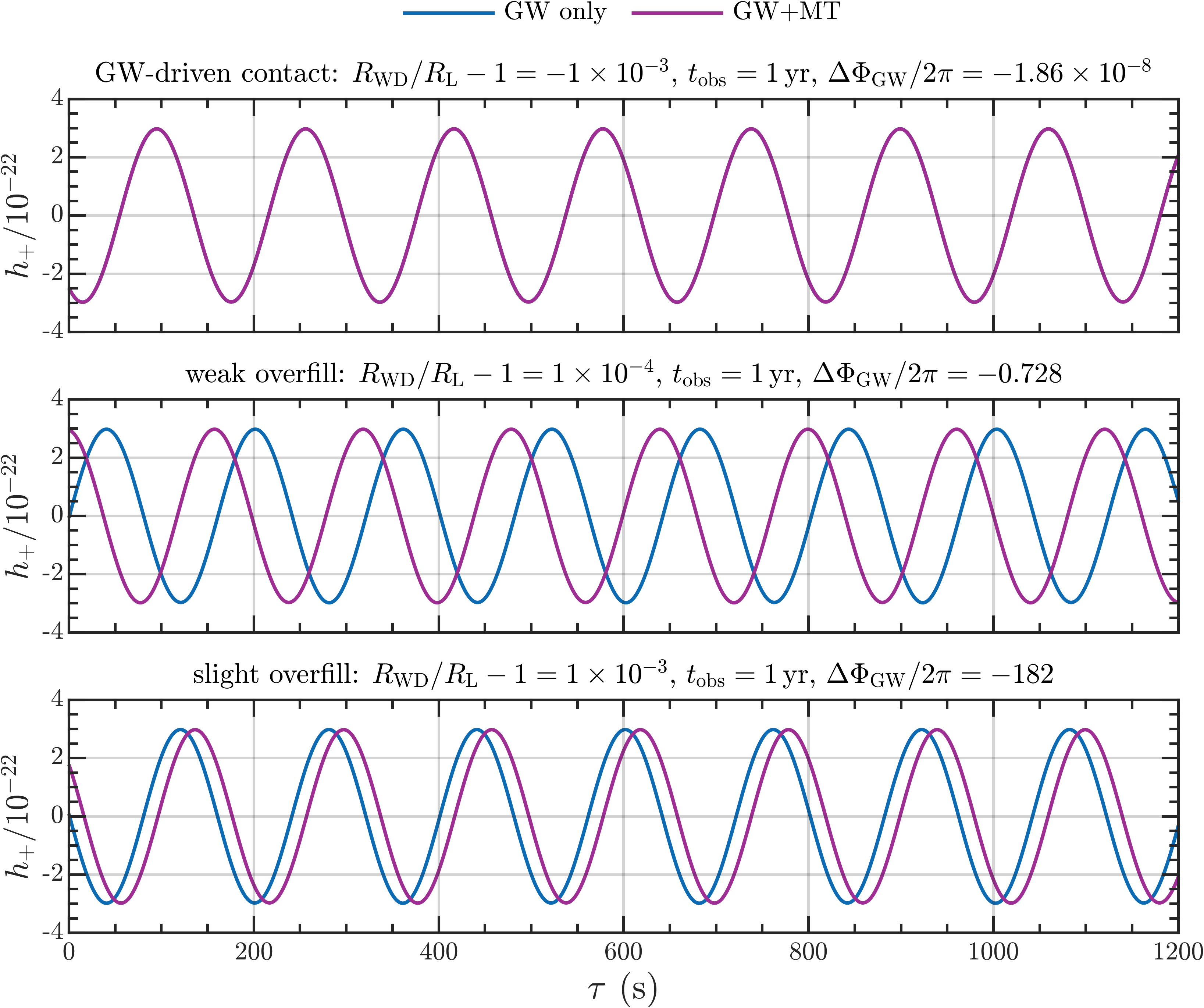}
		\includegraphics[width=0.48\textwidth]{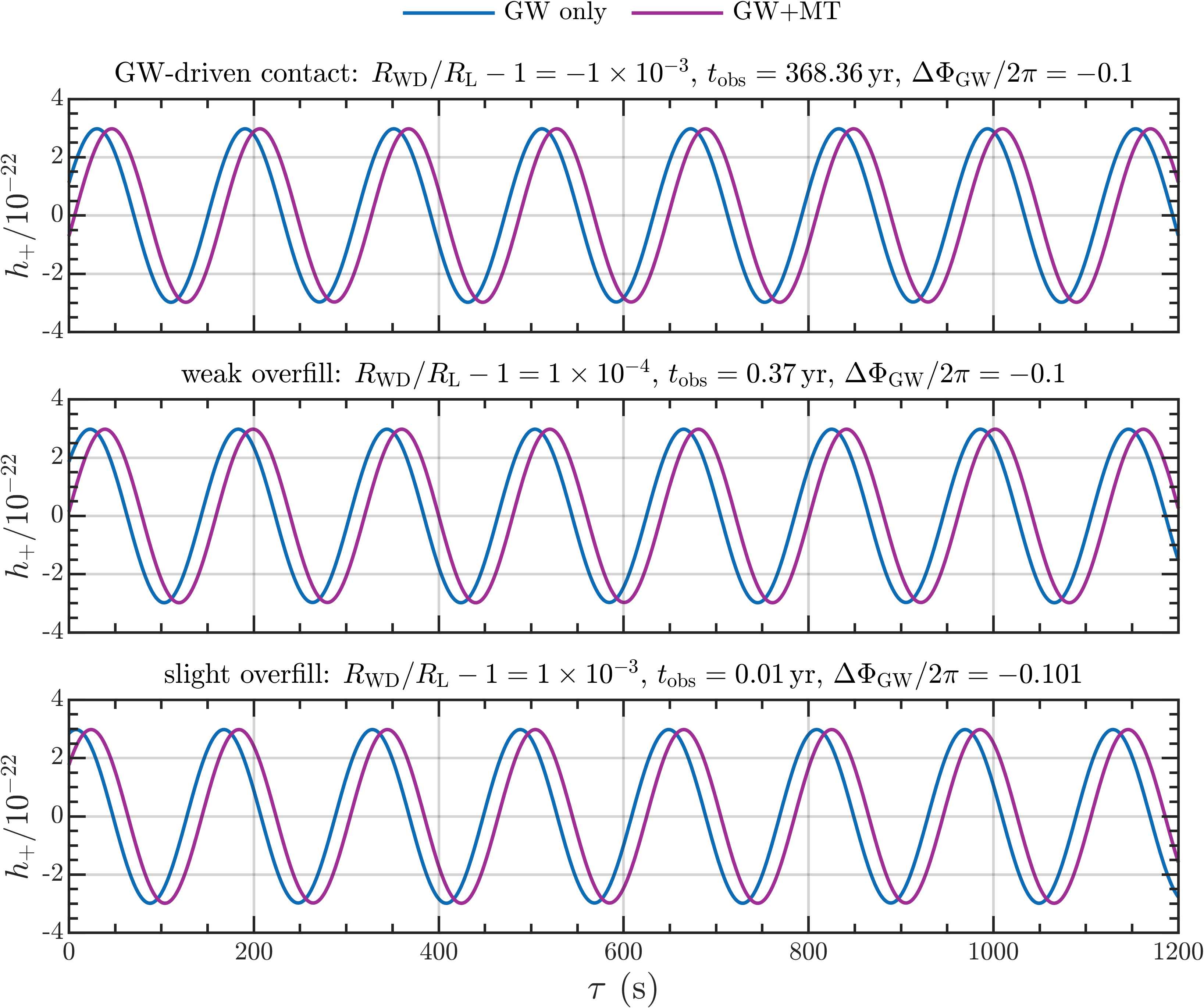}
		\caption{Time required for a visible waveform phase offset under different
			initial fill-out conditions. A larger initial
			overfill produces a larger mass-transfer rate and therefore a shorter
			timescale for phase separation.}
		\label{fig:phase-timescale}
	\end{figure*}
	
	Figure~\ref{fig:phase-timescale} compares the time needed for the mass-transfer
	phase correction to become visible in the waveform. In the detached,
	GW-driven-contact case, the phase separation develops only after a long
	evolution time. In the overfilled cases, the required time is much shorter.
	This behavior follows directly from the larger initial mass-transfer rate.
	The result also shows why the fill-out factor must be included when assessing
	whether mass transfer is observable in multi-year gravitational-wave data.
	The relevant detector context includes LISA, Taiji, and TianQin
	observations in the millihertz band \cite{Cutler1998,Cornish2003,Barack2004,AmaroSeoane2023,Hu2017,Luo2016,Robson2019,Huang2020}. 
	
	\clearpage
\section{Summary}

In this work we developed a first post-Newtonian description of Roche-lobe
mass transfer in compact binaries. A central result is the construction of the
1PN Roche potential from the 1PN hydrodynamic equations in the corotating
frame. After the terms depending on the fluid velocity are separated, the
remaining spacetime and non-inertial contributions define the effective Roche
potential at 1PN order. We also compared this definition with the relativistic
effective potential constructed from the helical Killing vector of a
conservative quasi-circular binary. The two constructions agree through 1PN
order and reduce to the Newtonian Roche potential in the Newtonian limit.

Using this 1PN Roche potential, we derived the corresponding 1PN
mass-transfer functions. For optically thin transfer, the isothermal Bernoulli
relation determines the density at \(L_1\), together with the
density-weighted cross section of the stream. For optically thick transfer,
the adiabatic streamline prescription gives the mass flux through the
equipotential area element near \(L_1\). Thus the mass-transfer rate is tied to
the local fluid dynamics and Roche geometry. This is different from approaches
in which the mass-transfer rate is prescribed externally. In our treatment,
the 1PN correction to the mass-transfer rate is obtained from the motion of
the transferred fluid itself.

We then introduced the resulting time-dependent masses into the binary
dynamics in the adiabatic mass-transfer regime. Keeping the first time
derivatives of the masses, we obtained the mass-transfer corrections to the
equations of motion and to the orbital-period evolution. We also calculated
the associated corrections to the gravitational-wave energy flux and angular
momentum flux. In the waveform, mass transfer enters through the slowly varying
mass parameters, the modified orbital phase, and explicit time derivatives of
the source multipole moments. These terms appear as mass-transfer corrections
to the far-zone polarization modes.

The numerical examples show that Roche-lobe overflow is self-regulated after
contact. In an initially overfilled, 4U 1820-30-like system, the mass-transfer
rate starts from a large value and rapidly relaxes toward a lower quasi-steady
state. In an initially underfilled, RX J0806.3+1527-like system,
gravitational radiation first drives the binary toward contact. After contact,
the mass-transfer rate rises and approaches the same regulated branch. The
period-derivative calculation shows that the mass-transfer contribution to
\(\dot P\) can exceed the Newtonian gravitational-wave contribution for the
mass-transfer rates considered here, while the 1PN correction to the
gravitational-wave term is much smaller.

We also applied the mass-transfer prescription to 4U 1728-34. This source is
not assumed to be a confirmed ultracompact X-ray binary. Instead, we used the
accretion luminosity to estimate the required mass-transfer rate and tested
whether a neutron-star--white-dwarf configuration can satisfy this
requirement. Our result shows that the luminosity-inferred mass-transfer range
can be reproduced with a small Roche-lobe overfill. This does not prove that
4U 1728-34 is an ultracompact X-ray binary, but it leaves this interpretation
open from the mass-transfer requirement alone.

Finally, we studied the effect of mass transfer on the gravitational waveform.
For the examples considered here, the instantaneous amplitude correction is
small. The dominant observable effect is instead a secular phase shift
accumulated over many orbital cycles. This accumulated phase can partially
cancel the gravitational-wave inspiral phase, and in some mass-transferring
systems it can dominate the net phase drift. This effect is relevant for
ultracompact binaries in the millihertz band, where space-based detectors may
observe the source over multi-year timescales \cite{AmaroSeoane2023,Robson2019,Kupfer2018}.

Several extensions are left for future work. The present calculation assumes a
simplified conservative mass-transfer model, an idealized donor structure, and
mostly circular binaries at leading 1PN order. More realistic stellar models,
thermal response, non-conservative mass loss, eccentricity, spin, tidal
coupling, and magnetic braking should be included in future studies. The local
hydrodynamic treatment near \(L_1\) should also be tested against numerical
simulations. A systematic comparison with observed ultracompact X-ray binaries
and space-based gravitational-wave verification sources will be needed to
assess whether mass-transfer-induced phase shifts can be measured in future
LISA, Taiji, or TianQin data \cite{AmaroSeoane2023,Hu2017,Luo2016}.
	
	\section{ACKNOWLEDGMENTS}
	This work is supported by the National Key Research and Development Program of China (No. 2021YFC2203003), 
	the National Natural Science Foundation of China (Grant Nos. 12247101) , and
	the “111 Center” (Grant No. B20063). SY is supported by the CAS Project for Young Scientists in Basic Research, Grant No. YSBR-063, and the National Natural Science Foundation of China (Grant No. 11673031).
	
	\appendix
	\begin{widetext}
	\section{Effective cross sections at the inner Lagrangian point 	\label{Effective cross sections}}

	In this appendix, we summarize the local construction of the effective cross
	sections near the inner Lagrangian point \(L_1\). The same \(L_1\)-plane
	geometry enters both the optically thin and optically thick prescriptions.
	
	We denote the effective Roche potential, including the 1PN correction, by
	\begin{equation}
		\phi_{\rm eff} = \phi_N+\frac{1}{c^2}\phi_{PN} ,
	\end{equation}
	where
	\begin{equation}
		\phi_N=-V -\frac{1}{2}\omega^2r^2 ,
		\qquad
		r^2=x^2+y^2 ,
	\end{equation}
	and
	\begin{equation}
		\phi_{PN} = -\frac{1}{4}\omega^4r^4 -2V\omega^2r^2 +4\omega \mathcal V_\phi ,
		\qquad
		\mathcal V_\phi=-yV_x+xV_y .
	\end{equation}
	Here \(r\) is the cylindrical radius measured from the rotation axis, and \(V_i=(V_x,V_y,V_z)\) is the PN vector potential.
	
	The 1PN-corrected \(L_1\) point is defined by
	\begin{equation}
		\nabla\phi_{\rm eff}\big|_{L_1}=0 .
	\end{equation}
	Writing
	\begin{equation}
		\mathbf r_{L_1} = \mathbf r_{L_1}^{(0)} + \frac{1}{c^2}\delta\mathbf r_{L_1},
	\end{equation}
	we find, to 1PN order,
	\begin{equation}
		\delta\mathbf r_{L_1}
		=- \left(H_0\right)^{-1}
		\left.\nabla\phi_1\right|_{\mathbf r_{L_1}^{(0)}} ,
		\qquad
		(H_0)_{ij} = \left.\partial_i\partial_j\phi_0
		\right|_{\mathbf r_{L_1}^{(0)}} .
	\end{equation}
	
	Near \(L_1\), we introduce local coordinates such that \(X\) is the direction
	of the stream through \(L_1\), while \(\xi^a=(y,z)\) span the transverse
	\(L_1\)-plane. The transverse expansion of the effective potential is
	\begin{equation}
		\phi_{\rm eff} = \phi_{L_1} + \frac{1}{2} H^{\rm PN}_{ab}\xi^a\xi^b +\cdots ,
	\end{equation}
	where
	\begin{equation}
		H^{\rm PN}_{ab} = \left. \frac{\partial^2\phi_{\rm eff}} {\partial \xi^a\partial \xi^b} \right|_{L_1}.
	\end{equation}
	This transverse Hessian includes both the explicit 1PN correction to the
	potential and the shift of the \(L_1\) point. Expanding
	\begin{equation}
		H^{\rm PN}_{ab} = H^{(0)}_{ab} + \frac{1}{c^2}H^{(1)}_{ab},
	\end{equation}
	gives
	\begin{equation}
		H^{(1)}_{ab} = \left. \partial_a\partial_b\Phi_1 \right|_{L_1^{(0)}} + \delta r_{L_1}^i \left. \partial_i\partial_a\partial_b\Phi_0 \right|_{L_1^{(0)}} .
	\end{equation}
	
	\subsection{Optically thin stream}
	
	In the optically thin prescription, the flow is treated as isothermal near
	\(L_1\). The isothermal sound speed \(c_T\) is fixed by the adopted equation
	of state and is not expanded in powers of \(c^{-2}\). This follows the
	standard optically thin RLO treatment \cite{Ritter1988,Meyer1983,Cehula2023}. The transverse density
	profile in the \(L_1\)-plane is then
	\begin{equation}
		\rho(\xi^a)
		=
		\rho_1
		\exp\left[
		-
		\frac{
			H^{\rm PN}_{ab}\xi^a\xi^b
		}{
			2c_T^2
		}
		\right] ,
	\end{equation}
	where \(\rho_1\) is the density at the center of the stream.
	
	We define the density-weighted effective cross section by
	\begin{equation}
		Q_\rho
		=
		\int_{L_1{\rm -plane}}
		\frac{\rho(\xi^a)}{\rho_1}\,d^2\xi .
	\end{equation}
	The Gaussian integral yields
	\begin{equation}
		Q_\rho^{\rm PN} = \frac{2\pi c_T^2} {\sqrt{\det H_\perp^{\rm PN}}} ,
	\end{equation}
	where \(H_\perp^{\rm PN}\) denotes the transverse \(2\times2\) matrix
	\(H_{ab}^{\rm PN}\). If this matrix is diagonal,
	\begin{equation}
		H_\perp^{\rm PN} =
		\begin{pmatrix}
			B_{\rm PN} & 0\\
			0 & C_{\rm PN}
		\end{pmatrix},
	\end{equation}
	then
	\begin{equation}
		Q_\rho^{\rm PN} = \frac{2\pi c_T^2} {\sqrt{B_{\rm PN}C_{\rm PN}}}.
	\end{equation}
	The optically thin mass-transfer rate is therefore
	\begin{equation}
		-\dot m_{\rm thin}^{\rm PN} = \rho_1 c_T Q_\rho^{\rm PN}.
	\end{equation}
	
	\subsection{Optically thick stream}
	
	In the optically thick prescription, the stream is treated as adiabatic \cite{Kolb1990}. The
	mass-transfer rate is obtained by integrating the mass flux over all
	streamlines that cross the \(L_1\)-plane,
	\begin{equation}
		-\dot m_{\rm thick} = \int_{L_1{\rm -plane}} \rho_L c_s\,dQ ,
	\end{equation}
	where \(\rho_L\) and \(c_s\) are evaluated on each streamline at the
	\(L_1\)-plane.
	
	In contrast to the optically thin case, the relevant geometrical factor is not
	a density-weighted area. Instead, it is determined by the area enclosed by an
	equipotential curve in the \(L_1\)-plane. Since
	\begin{equation}
		\Delta\phi \equiv \phi_{\rm eff}-\phi_{L_1}
		= \frac{1}{2}H^{\rm PN}_{ab}\xi^a\xi^b ,
	\end{equation}
	the intersection of an equipotential surface with the \(L_1\)-plane is an
	ellipse. Its area is
	\begin{equation}
		Q(\Delta\phi)
		=
		\frac{2\pi\Delta\phi}
		{\sqrt{\det H_\perp^{\rm PN}}}.
	\end{equation}
	Thus
	\begin{equation}
		\left.
		\frac{dQ}{d\phi}
		\right|_{L_1}
		=
		\frac{2\pi}
		{\sqrt{\det H_\perp^{\rm PN}}}.
	\end{equation}
	
	The optically thick mass-transfer rate can then be written as
		\begin{align}
		-\dot m_{\rm thick}
		&= -\dot m_{\rm thin,0} 
		+{2\pi\over \sqrt{B_{PN}C_{PN}}}
		\int_{\phi_1}^{\phi_{\rm ph}}
		F_3(\Gamma)
		\left( {k_B\bar T\over \bar m} \right)^{1/2}
		\bar\rho(\bar\phi)
		d\bar\phi  \\ 
		& = -\dot m_{\rm thin,0} +{2\pi\over \sqrt{B_{PN}C_{PN}}} F_3(\Gamma) K^{1/2} \int_{P_{ph}}^{P_{1}} P^{ \frac{\Gamma-1}{2 \Gamma}} dP \\
		& =-\dot m_{\rm thin,0}  +{2\pi\over \sqrt{B_{PN}C_{PN}}} F_3(\Gamma) K^{1/2}  P^{ \frac{3\Gamma-1}{2 \Gamma}} \bigg|^{P_{ph}}_{P_1}.
	\end{align}
	Here \(\dot m_{\rm thin,0}^{\rm PN}\) denotes the saturated contribution from
	the optically thin part of the atmosphere. The integral gives the contribution
	from the optically thick layers below the photosphere.
	
	Thus, the optically thin and optically thick prescriptions use the same local
	\(L_1\)-plane geometry, encoded in \(H_\perp^{\rm PN}\). They differ only in
	how this geometry enters the mass flux. The optically thin prescription uses
	the density-weighted area \(Q_\rho^{\rm PN}\), whereas the optically thick
	prescription uses the equipotential area element \(dQ/d\Phi\).
	\section{Multipole expansion of the radiation field \label{Multipole}}
	In this appendix, we summarize the multipole expansion used to compute the
	gravitational-wave energy and angular-momentum fluxes. We follow the standard
	symmetric trace-free (STF) conventions \cite{Thorne1980,Will1996,Ross2012,Blanchet2024}. For a multi-index
	\(L=i_1\cdots i_\ell\), the source moments of the effective stress-energy
	pseudo-tensor are written as
	\begin{align}
		\text{F}_{\mu\nu L}&=\underset{B=0}{\text{PF}}\int \mathrm{d}^3 x \Big(\frac{r}{r_0}\Big)^B\hat{x}_L\int^1_{-1} \mathrm{d} z\delta_\ell (z)\tau^{\mu\nu}(\bm{x},t+zr/c),
	\end{align}
	where \(\underset{B=0}{\text{PF}}\) denotes the finite part at \(B=0\).
	The factor \((r/r_0)^B\) regularizes the near-zone integral. The integration
	over \(z\) is weighted by
	\begin{equation}
		\delta_\ell(z)=\frac{(2\ell+1)!!}{2^{\ell+1}\ell !}(1-z^2)^\ell,
	\end{equation}
	The far-zone metric perturbation is then
	\begin{align}
		h^{\mu\nu}=-\frac{4G}{c^4}\sum_{\ell\geq0}\frac{(-)^\ell}{\ell !}\partial_L\bigg[\frac{1}{r}F^{\mu\nu}_L(t-r/c)\bigg].
	\end{align}
	The gravitational-wave energy and angular-momentum fluxes are
	\begin{align}
		\mathcal{F}_{GW}=&\sum^{+\infty}_{\ell=2}\frac{G}{c^{2\ell+1}}\frac{(\ell+1)(\ell+2)}{(\ell-1)\ell\ell!(2\ell+1)!!}\bigg[\text{U}^{(1)}_{L}\text{U}^{(1)}_{L}+\frac{4\ell^2}{c^2(\ell+1)^2}\text{V}^{(1)}_L\text{V}^{(1)}_L\bigg],\\
		\mathcal{N}^i_{GW}=&\sum^{+\infty}_{\ell=2}\frac{G}{c^{2\ell+1}}\frac{(\ell+1)(\ell+2)}{(\ell-1)\ell!(2\ell+1)!!}\bigg[\text{U}_{jL-1}\text{U}^{(1)}_{kL-1}+\frac{4\ell^2}{c^2(\ell+1)^2}\text{V}_{jL-1}\text{V}^{(1)}_{kL-1}\bigg]\epsilon_{ijk}.
	\end{align}
	At leading order in the post-Minkowskian matching, the radiative moments are
	related to the source moments by
	\begin{equation}
		\mathrm{U}_L=\mathrm{I}^{(\ell)}_L+\mathcal{O}(G),\quad \mathrm{V}_L=\mathrm{J}^{(\ell)}_L+\mathcal{O}(G).
	\end{equation}
	Keeping the terms required for 1PN accuracy gives
	\begin{align}
		\mathcal{F}_{GW}=&\frac{G}{c^5}\frac{1}{5}\big\langle \hat{\mathrm{I}}^{(3)}_{ij}\hat{\mathrm{I}}^{(3)}_{ij}\big\rangle+\frac{G}{c^7}\bigg[\frac{16}{45}\big\langle \hat{\mathrm{J}}^{(3)}_{ij}\hat{\mathrm{J}}^{(3)}_{ij}\big\rangle+\frac{1}{189}\big\langle \hat{\mathrm{I}}^{(4)}_{ijk}\hat{\mathrm{I}}^{(4)}_{ijk}\big\rangle\bigg]+\mathcal{O}(c^{-9}),\\
		\mathcal{N}^i_{GW}=&\frac{G}{c^5}\frac{2}{5}\big\langle \hat{\mathrm{I}}^{(2)}_{jl}\hat{\mathrm{I}}^{(3)}_{kl}\big\rangle\epsilon_{ijk}+\frac{G}{c^7}\bigg[\frac{32}{45}\big\langle \hat{\mathrm{J}}^{(2)}_{jl}\hat{\mathrm{J}}^{(3)}_{kl}\big\rangle+\frac{1}{63}\big\langle \hat{\mathrm{I}}^{(3)}_{jlm}\hat{\mathrm{I}}^{(4)}_{klm}\big\rangle\bigg]\epsilon_{ijk}+\mathcal{O}(c^{-9}),
	\end{align}
	Here \(\mathrm{I}_{ij}\), \(\mathrm{I}_{ijk}\), and \(\mathrm{J}_{ij}\)
	are the mass quadrupole, mass octupole, and current quadrupole, respectively.
	A hat denotes the STF projection. For example,
	\begin{gather}
		\hat{\mathrm{A}}_{ij}={\mathrm{A}}_{\langle ij\rangle}=\mathrm{A}_{(ij)}-\frac{1}{3}\delta_{ij}\mathrm{A}_{kk},\quad \hat{\mathrm{A}}_{ijk}={\mathrm{A}}_{\langle ijk\rangle}=\mathrm{A}_{(ijk)}-\frac{1}{5}\Big(\delta_{ij}\mathrm{A}_{llk}+\delta_{ik}\mathrm{A}_{ljl}+\delta_{jk}\mathrm{A}_{ill}\Big).
	\end{gather}
	
	The mass quadrupole is required through 1PN order. We use the standard source
	multipole expression \cite{Will1996,Blanchet2024,Henry2024}
	\begin{align}
		\hat{\mathrm{I}}_{jk}=&\Big(\mu x_j x_k-\frac{1}{3}\mu\delta_{jk} r^2\Big)\bigg\{1+\frac{1}{c^2}\Big(\frac{a}{\mathcal{Z}} v^2-b \frac{Gm}{r}\Big)\bigg\}+\mathcal{O}(c^{-4}),\\
		\hat{\mathrm{I}}^{ij}=&\int \mathrm{d}^3\mathbf{x} \sigma \hat{x}^{ij}+\frac{1}{c^2}\frac{1}{14}\frac{\mathrm{d}^2}{\mathrm{d}t^2}\int \mathrm{d}^3\mathbf{x} |x|^2\sigma\hat{x}^{ij}-\frac{1}{c^2}\frac{20}{21}\frac{\mathrm{d}}{\mathrm{d}t}\int\mathrm{d}^3\mathbf{x}\sigma_k \hat{x}^{kij}+\mathcal{O}(c^{-4}),
	\end{align}
	The mass octupole and current quadrupole are needed only at Newtonian order:
	\begin{align}
		\hat{\mathrm{I}}^{ijk}=\int \mathrm{d}^3\mathbf{x} \sigma \hat{x}^{ijk}+\mathcal{O}(c^{-4}),\quad \hat{\mathrm{J}}^{ij}=\int \mathrm{d}^3\mathbf{x} \epsilon_{ijk}x_{i}\sigma_j \hat{x}^{ij}+\mathcal{O}(c^{-4}),
	\end{align}
	In reducing the time derivatives, we use the following STF identities:
	\begin{gather}
		x^{\langle i}v^{k\rangle}v^k=\frac{1}{2}v^2x^{i}-\frac{1}{6}r\dot{r}v^i,\quad x^{\langle ik \rangle}x^k=\frac{2}{3}r^2x^i,\\
		x^{\langle ij}v^{k\rangle}v^k=\frac{1}{3}x^{\langle ij \rangle}v^2+\frac{2}{5}r^{\langle i}v^{j\rangle}r\dot{r}-\frac{2}{15}v^{\langle ij \rangle}r^2,\quad x^{\langle ijk \rangle}x^k=\frac{3}{5}x^{\langle ij\rangle}r^2.
	\end{gather}
	
	The required time derivatives of the mass quadrupole are
	\begin{align}\label{DI}
		\begin{split}
			\hat{\mathrm{I}}^{(1)}_{jk} &= 2\mu P^{ik} + \dot{\mu} Q_x^{ik}  \\
			&+\frac{\mu}{c^2} \biggl\{ \biggl[ \frac{17a}{21}v^2 -\frac{2Gm}{21r}(7+13a) \biggr] P^{ik} +\frac{10a}{21}r\dot r\,Q_v^{ik} +\frac{Gm}{21r^2}(7-9a)\dot r\,Q_x^{ik} \biggr\}  \\
			&+\frac{\dot\mu}{c^2} \biggl\{
			\frac{4a}{7}r\dot r\,P^{ik} -\frac{11a}{21}r^2Q_v^{ik}
			+\biggl[ \frac{29a}{42}(v^2-2v\dot r)  -\frac{Gm}{21r}(7+8a) \biggr] Q_x^{ik} \biggr\} +\mathcal O(c^{-4}) . 
		\end{split}\\
		\begin{split}
			\hat{\mathrm{I}}_{jk}^{(2)} &= 2\mu Q_v^{ik} -\frac{2Gm\mu}{r^3}Q_x^{ik} + 2 \dot{\mu}P^{ik} \\
			&+\frac{\mu}{c^2} \biggl\{
			Q_v^{ik}\biggl[ \frac{9a}{7}v^2 -\frac{2Gm}{21r}(7+18a) \biggr] +\frac{2Gm}{7r^2}(28-3a)\dot r\,P^{ik} \\
			&\qquad+Q_x^{ik}\biggl[
			\frac{G^2m^2}{3r^4}(29+a)
			+\frac{Gm}{21r^3}\bigl((16a-77)v^2+6a\dot r^2\bigr)\biggr]\biggr\} \\
			&+ \frac{\dot\mu }{c^2}  
			\biggl\{ -\frac{10a}{7}r\dot r\,Q_v^{ik} 
			+P^{ik}\biggl[ \frac{a}{21}(41v^2-92v\dot r)-\frac{2Gm}{21r}(7+3a) \biggr] \\
			&\qquad +Q_x^{ik}\biggl[ \frac{29a}{21r}v(\dot r^2-v^2) +\frac{Gm}{21r^2}
			\biggl(29a\biggl(v+\frac{\dot r^2}{v}\biggr) +(7-33a)\dot r \biggr) 	\biggr] \biggr\} +\mathcal O(c^{-4}) . 
		\end{split} \\
		\begin{split}
			\hat{\mathrm{I}}^{(3)}_{jk}
			&= -\frac{8Gm\mu}{r^3}P^{jk}  -\frac{4Gm}{r^3} \dot{\mu}Q_x^{jk} \\
			&+ \frac{\mu}{c^2} \biggl\{ \frac{2G^2m^2}{21 r^4} (329+24a)- \frac{2Gm}{21r^4} 11(7-2a)rv^2 \biggr\}P^{jk}\\
			&+\frac{\dot\mu}{c^2}
			\biggl\{ -\frac{16a}{21}v^2Q_v^{jk}+\frac{10a}{7r}v^j(2Gm v^k-5v^3x^k) \\
			&\qquad+\frac{2Gm}{21r^4} \biggl[ Gm(203+9a)-(77+8a)rv^2 \biggr]Q_x^{jk}
			+\frac{50Gma}{7r^2}vP^{jk} \biggr\} +\mathcal O(c^{-4}) .
		\end{split}
	\end{align}
	The Newtonian mass octupole and its time derivatives are
	\begin{align}
		\mathrm{I}_{ijk}&=\mu x_ix_j x_k-\frac{3}{5}r^2\delta_{(ij}x_{k)}+\mathcal{O}(c^{-2}),\\
		{\mathrm I}^{(1)}_{ijk}
		&= \mu\biggl[-\frac35 r^2\delta^{(ij}v^{k)}-v^ix^jx^k-\frac65 r\dot r\,\delta^{(ij}x^{k)}\biggr]-\frac35 r^2\dot\mu\,\delta^{(ij}x^{k)}+\mathcal O(c^{-2}),\\
		{\mathrm I}^{(2)}_{ijk}
		&=\mu\biggl[-2v^iv^kx^j-\frac65 r\biggl(-\frac{Gm}{r^2}+\frac{v^2}{r}\biggr)\delta^{(ij}x^{k)}+2x^kx^{(i}\mathcal A^{j)}-\frac35 r^2\delta^{(ij}\mathcal A^{k)}\biggr]+2\dot\mu\biggl[-\frac35 r^2\delta^{(ij}v^{k)}-v^ix^jx^k\biggr]+\mathcal O(c^{-2}),\\
		{\mathrm I}^{(3)}_{ijk}
		&= \frac15\biggl[3\mu\biggl(\frac{7Gm}{r}-6v^2\biggr)\delta^{(ij}v^{k)}+\frac{35Gm\mu}{r^3}v^ix^jx^k+3\dot\mu\biggl(\frac{6Gm}{r}-4v^2\biggr)\delta^{(ij}x^{k)}\biggr]+\mathcal O(c^{-2}),\\
		\hat{\mathrm I}^{(4)}_{ijk}
		&= \frac35\biggl[\frac{Gm\mu}{r^4}\bigl(-7Gm+6rv^2\bigr)\delta^{(ij}x^{k)}+\dot\mu\biggl(\frac{6Gm}{r}-4v^2\biggr)\delta^{(ij}v^{k)}\biggr]+\mathcal O(c^{-2}) ,
	\end{align}
	where \(\mathcal A^i=-Gm x^i/r^3-v^i\dot\mu/\mu\). The required current-quadrupole moments are
	\begin{align}
		\mathrm{J}_{jk}&=\mu x_{(k}\epsilon_{j)lm} x_l v_m +\mathcal{O}(c^{-2}),\\
		{\mathrm J}^{(1)}_{ij}
		&=\mu v^{(i}\epsilon^{j)lm}x^lv^m+\mathcal O(c^{-2}),\\
		{\mathrm J}^{(2)}_{ij}
		&=\frac{Gm\mu}{r^3}x^{(i}\epsilon^{j)lm}x^lv^m-\dot\mu v^{(i}\epsilon^{j)lm}x^lv^m+\mathcal O(c^{-2}),\\
		{\mathrm J}^{(3)}_{ij}
		&=\frac{Gm\mu}{r^3}v^{(i}\epsilon^{j)lm}x^lv^m
		-\biggl(\frac{3Gm\mu\dot r}{r^4}+\frac{Gm\dot\mu}{r^3}\biggr)x^{(i}\epsilon^{j)lm}x^lv^m
		+\mathcal O(c^{-2}) .
	\end{align}
	\end{widetext}

	\bibliography{reference}
	\bibliographystyle{apsrev4-1}
\end{document}